\documentclass[aps,pra,reprint,amsmath,amssymb,superscriptaddress]{revtex4-1}
\usepackage{amsmath}
\usepackage{times}
\usepackage[english]{babel}
\usepackage[caption = false]{subfig}
\usepackage[colorlinks=true,linkcolor=blue,urlcolor=blue,citecolor=blue]{hyperref}
\usepackage{color}
\usepackage{amscd}
\usepackage{amsfonts}
\usepackage{amssymb}
\usepackage{graphicx, epstopdf}
\usepackage{tabularx}
\usepackage{braket}
\usepackage{bm}
\usepackage{dsfont}

\newcommand{\teqref}[1]{Eq.~\eqref{#1}}
\newcommand{\teqsref}[2]{Eqs.~\eqref{#1}-\eqref{#2}}

\newcommand{\tcite}[1]{~\cite{#1}}
\newcommand{\tref}[1]{~\ref{#1}}
\newcommand{\ov}{``}

\newlength\mytemplen
\newsavebox\mytempbox

\makeatletter
\newcommand\mybluebox{%
    \@ifnextchar[%]
       {\@mybluebox}%
       {\@mybluebox[0pt]}}

\def\@mybluebox[#1]{%
    \@ifnextchar[%]
       {\@@mybluebox[#1]}%
       {\@@mybluebox[#1][0pt]}}

\def\@@mybluebox[#1][#2]#3{
    \sbox\mytempbox{#3}%
    \mytemplen\ht\mytempbox
    \advance\mytemplen #1\relax
    \ht\mytempbox\mytemplen
    \mytemplen\dp\mytempbox
    \advance\mytemplen #2\relax
    \dp\mytempbox\mytemplen
    \colorbox{myblue}{\hspace{1em}\usebox{\mytempbox}\hspace{1em}}}

\begin{document}

\title{Dephasing-enhanced performance in quasiperiodic thermal machines}

\author{Cecilia Chiaracane}
\email{chiaracc@tcd.ie}
\affiliation{School of Physics, Trinity College Dublin, Dublin 2, Ireland}
\author{Archak Purkayastha}
\email{archak.p@tcd.ie}
\affiliation{School of Physics, Trinity College Dublin, Dublin 2, Ireland}
\author{Mark T. Mitchison}
\email{mark.mitchison@tcd.ie}
\affiliation{School of Physics, Trinity College Dublin, Dublin 2, Ireland}
\author{John Goold}
\email{gooldj@tcd.ie}
\affiliation{School of Physics, Trinity College Dublin, Dublin 2, Ireland}

\begin{abstract}
Understanding and controlling quantum transport in low-dimensional systems is pivotal for heat management at the nanoscale. One promising strategy to obtain the desired transport properties is to engineer particular spectral structures. In this work we are interested in quasiperiodic disorder --- incommensurate with the underlying periodicity of the lattice --- which induces fractality in the energy spectrum. A well known example is the Fibonacci model which, despite being non-interacting, yields anomalous diffusion with a continuously varying dynamical exponent smoothly crossing over from superdiffusive to subdiffusive regime as a function of potential strength.  We study the  finite-temperature electric and heat transport of this model in linear response in the absence and in the presence of dephasing noise due to inelastic scattering. The dephasing causes both thermal and electric transport to become diffusive, thereby making thermal and electrical conductivities finite in the thermodynamic limit. Thus, in the subdiffusive regime it leads to enhancement of transport. We find that the thermal and electric conductivities have multiple peaks as a function of dephasing strength. Remarkably, we observe that the thermal and electrical conductivities are not proportional to each other, a clear violation of Wiedemann-Franz law, and the position of their maxima can differ. We argue that this feature can be utilized to enhance performance of quantum thermal machines. In particular, we show that by tuning the strength of the dephasing noise we can enhance the performance of the device in regimes where it acts as an autonomous refrigerator.
\end{abstract}

\date{\today}
\maketitle

\section{Introduction}
The progressive miniaturization of technology has boosted the search for quantum devices beyond semiconductors that would improve the micromanagement of heat in solid state devices \tcite{whitney2014most}.
The premise on which most solid state physics is based is the notion of periodicity. This gives rise to a band structure and, due to translational symmetry, extended single particle states known as Bloch waves. This picture is modified due to the inevitable presence of disorder. In mesoscopic physics the interplay between transport and disorder is well studied and in particular this interplay has been shown to enhance the thermoelectric performances of disordered low dimensional systems\tcite{sivan1986,hicks1, hicks2,bosisio2014, bosisio2014B, sanchez2013, muttalib2015, linke2010,Boukai2008, Hochbaum2008,  curtin2014,yamamoto2017,dominguez, chiaracane}. 

A special type of disorder is represented by quasiperiodic potentials, incommensurate with the underlying periodicity of the  lattice\tcite{simon1982, ostlund1983, ostlund1984}. These systems, often called quasicrystals, are known to possess highly non-trivial singular continuous spectra with fractal structure\tcite{Bellissard1989}, which leads to the appearance of critical states\tcite{Kohmoto1987, hiramoto} which are neither extended nor localized. These unique spectral features can in fact induce localization and anomalous transport without the presence of interactions\tcite{jagannathan2021, varma2016, piechon1996, vkv2017, Fibotransport, nori1986, Kohmoto1983}. Perhaps the most celebrated example is the Aubry-Andr\'e-Harper (AAH) model\tcite{aubry1980analyticity,harper} which displays a transition from a completely delocalized to a completely localized phase at a finite  potential strength. At the critical point the transport is known to be anomalous\tcite{purkayastha2018}. The AAH model has a wide range of generalizations~\cite{GAAH1, GAAH2, GAAH3, GAAH4, dasSarma_bichromatic}, where the localization transition can become energy-dependent due to a mobility edge. A closely related model, which is topologically connected to the AAH model \tcite{Goblot2020, zilberberg2012}, is the Fibonacci model where the lattice energies are generated by a substitution rule. The Fibonacci quasicrystal has unusual properties such as a critical energy spectrum across all energy scales\tcite{jagannathan2021, mace2016, hiramoto1988, Zhong_1995}, without a localization transition. This spectral criticality gives rise to anomalous transport exponents varying continuously with the potential strength, so that it is possible to tune the transport regime from superdiffusive to subdiffusive\tcite{hiramoto1988, Fibotransport, varma2019, chiaracane2,lacerda2021}. 

Quasiperiodic quantum systems and their spectra have been intensely studied in pure mathematics\tcite{simon1982, Last1994}, but are also relevant for a strikingly diverse range of physical systems. Quasiperiodic models have also been shown to offer potential applications in quantum heat management, e.g.~as rectifiers\tcite{balachandran2019, saha2019, saha2017} or as highly efficient working media for thermoelectric engines\tcite{chiaracane}. Beside being recently identified in compounds found in meteors\tcite{Bindi2009, Bindi2015}, they arise in experiments with ultra-cold atomic gases\tcite{exp2, expmbloc,expmbloc3,expmbloc4}, and photonics\tcite{Goblot2020,dalnegro2003,Man2005, lahini2012, verbin2015}, where the effective potential is modulated to be quasiperiodic by tuning respectively the wavevectors of overlapping optical lattices and the refraction indices of coupled waveguide arrays. Moreover, quasiperiodic arrangements of nucleotides in synthetic DNA molecules have been proposed to realize nanoelectric devices\tcite{macia2006, guo2012}. In single DNA molecules, transport is characterized by a concurrence of coherent and incoherent mechanisms, determined by the interaction between conducting electrons and \ov environmental'' degrees of freedom such as the other electrons, nuclei or the solvent\tcite{korol2016, kim2017, Xiang2015, Bruot2015}. These many-body effects collectively introduce noise that might consist of loss of phase coherence, and momentum and energy exchange. It has been demonstrated in various contexts that this noise from the environment can assist transport. The examples of such environmental assisted or dephasing enhanced transport include natural photosynthetic complexes \tcite{Plenio2008, rebentrost2009, caruso2009, Chin2012, collini, scholes, zerah2018}, molecular junctions\tcite{kilgour2015B, kilgour2016B,sowa2017}, photonic crystals\tcite{Biggerstaff2016, Viciani2016, Caruso2016}, trapped ions\tcite{zerah2020, trautmann2018}, and also boundary-driven spin chains at infinite temperature\tcite{znidaric2013,znidaric2017,lacerda2021}.
However, the implications of this effect for thermoelectricity --- an intrinsically finite-temperature phenomenon --- have received comparatively little attention.  Here we ask if the inevitable presence of dephasing noise due to inelastic scattering can be used to enhance thermoelectric performance of quasicrystals.

In particular, we investigate steady-state thermoelectric transport in the Fibonacci model in presence of both temperature and chemical potential bias. We find that the anomalous transport behavior observed previously at infinite temperature survives at finite temperatures in both the electric and thermal transport. However, noise in the form of incoherent inelastic scattering, leading to dephasing and energy relaxation, causes the system to lose the anomalous behavior by making transport diffusive. We study the electric and thermal conductivities, well-defined and finite only for diffusive transport, as a function of the dephasing strength. We demonstrate that the conductivities can be enhanced by bulk incoherent effect including phase loss and energy exchange in  the subdiffusive regime of the Fibonacci model at finite temperature. Interestingly, we find that the optimal dephasing strength may be markedly different for the electric and thermal conductivities. This constitutes a clear violation of Wiedemann-Franz law, which says the thermal and the electric conductivities at a given temperature are proportional to each other. In fact, we find that the Wiedemann-Franz law is violated for a wide range of dephasing strengths, despite the transport being diffusive. We argue that this dephasing-induced discrepancy between electric and heat transport can be exploited to improve performance of autonomous heat engines and refrigerators. In particular, we demonstrate that, in certain parameter regimes, the dephasing noise from inelastic scattering can simultaneously enhance both the cooling rate and the coefficient of performance of an autonomous refrigerator with the Fibonacci quasicrystal as a working medium.

The outline of the paper is as follows. In Sec.\tref{sec:model}, we introduce the Fibonacci model. In Sec.\tref{sec:classification} we discuss how electric and thermal transport can be classified in presence of both temperature and chemical potential biases. In Sec.\tref{sec:coherent}, we investigate the anomalous transport properties of the Fibonacci model in the coherent regime, i.e, in absence of inelastic scattering. In Sec.\tref{sec:deph}, we explore the effect of incoherent inelastic scattering on electric and thermal transport properties in the framework of B{\"u}ttiker probes. In Sec.\tref{sec:frigo}, we discuss how the highly non-trivial transport properties of the Fibonacci model in presence of dephasing noise from inelastic scattering can be used to enhance refrigeration in the device in certain favorable thermodynamic configurations. Finally, we summarize and draw our conclusions in Sec.\tref{sec:concl}. 

\section{Fibonacci model}
\label{sec:model}
In this work we focus on a specific example from the family of quasiperiodic systems, the Fibonacci model\tcite{jagannathan2021, hiramoto1988}. We take a one-dimensional (1D) tight-binding chain of non-interacting fermions, described by the following Hamiltonian
\begin{equation}
\label{eq:Fwire}
\hat{H}_{F}=\sum_{n=1}^{N-1}  t (\hat{a}^{\dagger}_{n}\hat{a}_{n+1}+{\rm h.c})+ \sum_{n=1}^{N} u_n\hat{a}^{\dagger}_{n}\hat{a_{n}},
\end{equation}
with $t$ the tunnelling constant and $\hat{a}_n$ the fermionic annihilation operator of site $n$. The on-site energies $u_n$ are alternatively chosen between two values $(u_A, u_B)$ according to a Fibonacci substitution rule. The total collection of values $C_k = [u_1, u_2, ..., u_k]$ for a chain of size $F_k$ is obtained by iterating $k$ times the transformation
\begin{align}
        u_A & \rightarrow u_A u_B  \\
        u_B & \rightarrow u_A.
\end{align}
Equivalently, it can be generated by concatenation of two smaller chains $C_k = [C_{k-1}, C_{k-2}]$, starting from $C_0 = [ u_B], \ C_1 =  [u_A]$. As a consequence, the length of every chain $C_k$ is a number from the Fibonacci sequence $F_k \in \{1, 1,2,3,5, 8, \dots \}$. Particles in the model are subject to quasiperiodic disorder, which is deterministic and not random, but represents the closest example to periodicity\tcite{nori1986, hiramoto}. Quasiperiodic systems cannot be generated by repeating a smaller unit cell, yet in the indefinitely extended limit the frequency at which the same values of the potential occurs has a definite limit: in this example, the frequency of $u_B$ relative to $u_A$ becomes $\tau$ in the limit $k \rightarrow \infty$, with $\tau = (1+\sqrt{5})/2$ the golden ratio~\cite{goodson2017}. For this reason, quasiperiodic lattices are often considered as periodic systems with an infinite period\tcite{ostlund1983}.

Results for quasiperiodic systems are dependent on the choice of system sizes\tcite{purkayastha2018,vkv2017,Sutradhar2019}. For the Fibonacci potential in particular, they depend on how different $N$ is from a Fibonacci number. To reduce this dependence on choice of system sizes, we use the averaging procedure adopted in Refs.~\cite{varma2019,mace2016,chiaracane2}.  In order to treat arbitrary lengths $N$ which do not belong to the Fibonacci sequence, we cut finite samples of length $N$ out of a long Fibonacci potential sequence $C_{k}$, with $k$ such that $F_k \gg N$. After discarding the examples which are reflection-symmetric around the centre of the chain and their symmetric partners, there exist $N/2$ (or $(N-1)/2$ if $N$ is odd) samples with nonequivalent energy spectra available to average over. This averaging procedure also restores effective translational invariance in the thermodynamic limit. 

\section{Classification of transport in presence of both temperature and voltage biases}
\label{sec:classification}
In the linear response regime, electric  and heat  currents, $J_e$ and $J_q$ respectively, can be rewritten as linear combination of the driving biases~\cite{callen,Houten1992,de2013non,benenti2017},
\begin{align}
\label{eq:currcoff}
   J_e &= G \Delta \mu/e + G S \Delta T, \\
   \label{eq:currcoff2}
   J_q &= G \ \Pi \Delta \mu /e + (K + G S \ \Pi) \Delta T,
\end{align}
with $G$ and $K$ respectively the electric and the heat conductances, $S$ the thermopower or Seebeck coefficient, and $\Pi$ the Peltier coefficient, which in presence of time-reversal symmetry differs from $S$ only by a factor of $1/T$. The electric and the thermal conductivities are given by
\begin{align}
&    \sigma = \lim_{N\rightarrow \infty} \sigma(N),~~\sigma(N)= N G, \\
& \kappa =\lim_{N\rightarrow \infty} \kappa(N),~~\kappa(N)= N K.
\end{align}
If the system-size scaling of the conductances is
\begin{align}
    G \sim N^{-\alpha_G},~~K\sim N^{-\alpha_K},
\end{align}
we immediately see that the conductivities are well-defined and finite only if $\alpha_G=\alpha_K=1$. This corresponds to normal diffusive transport. Ballistic transport corresponds to the case $\alpha_G,\alpha_K=1$, whereas, if   $\alpha_G,\alpha_K<1$, the transport is called anomalous superdiffusive. In both these cases, $\sigma(N)$ and $\kappa(N)$ diverge with $N$, so the conductivities are ill-defined. On the other hand, if $\alpha_G,\alpha_K>1$, the conductivities are zero, and this corresponds to anomalous subdiffusive transport. If $G$, $K$ decay exponentially with system size, instead of a power-law, it signifies complete lack of transport. Thus, finite-size scaling of the conductances $G$ and $K$ can be used to characterize the nature of electric and heat transport.

\section{Anomalous coherent transport}
\label{sec:coherent}
We first reproduce this regime of anomalous transport in absence of dephasing within Landauer's framework for a two-terminal device. In the Fibonacci model, the quasiperiodicity of the potential induces a multi-fractal spectrum at every $u_A$ and $u_B$\tcite{Kohmoto1987, mace2016}, meaning that a self-similar structure emerges at different energy scales. Therefore, one may assume control over a single parameter $u_A = - u_B = u$ without loss of generality. It is known that multifractality yields anomalous behaviour in the Fibonacci model when transport is coherent: currents scale with system size as power laws $J \sim 1/N^{\alpha}$, where the exponent $\alpha$ varies continuously with $u$, from super-diffusive ($\alpha <1$) to sub-diffusive ($\alpha >1$) behaviour through normal diffusion ($\alpha = 1$)\tcite{vkv2017, hiramoto1988, piechon1996}. However, existing calculations focus on the infinite temperature case, and surprisingly, to our knowledge, the survival of this feature has not been demonstrated at finite temperatures. Moreover, the thermoelectric response of the Fibonacci model in the presence of both temperature and chemical potential biases has also not been explored before.

\begin{figure}[t]
\centering
\includegraphics[width=0.98\columnwidth]{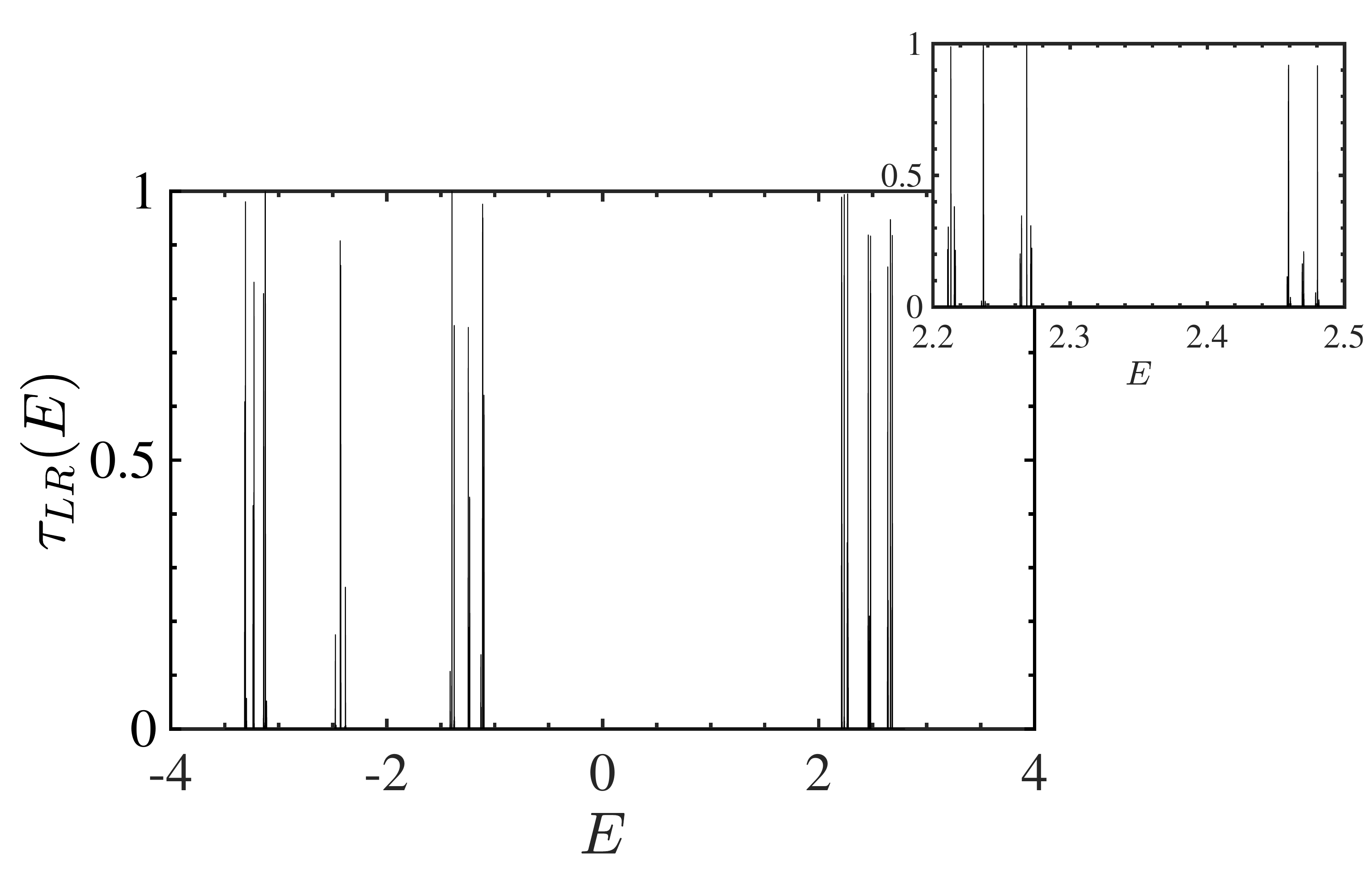}
\caption{Example of zero-dephasing transmission function $\tau_{LR}(E)$ of a single Fibonacci chain realization of size $N=200$, at $u=2.0$. In the inset, we explicitly show the self-similarity of the structure by zooming on a portion of the energy axis.}
\label{fig:transmission}
\end{figure}

We consider a region of elastic scattering governed by the Fibonacci Hamiltonian, which is connected at its boundaries to metallic leads that are initially in thermal equilibrium at temperatures $T_L$, $T_R$, and chemical potentials $\mu_L$, $\mu_R$. The total Hamiltonian is 
\begin{equation}
\label{eq:H}
\hat{H} = \hat{H}_F + \sum_{\nu} ( \hat{H}_{\nu} + \hat{H}_{F \nu}),
\end{equation} 
where $\hat{H}_F$ is given in \teqref{eq:Fwire}, and
\begin{equation}
\label{eq:H_bath}
\hat{H}_{\nu}=\sum_{\lambda}E_{\lambda\nu}\hat{d}_{\lambda\nu}^{\dagger}\hat{d}_{\lambda\nu},
\end{equation}
with $E_{\lambda\nu}$ the single-particle eigenenergies of lead $\nu = L,R$ and $\hat{d}_{\lambda\nu}$ the annihilation operators for the corresponding eigenmodes $\lambda$. We connect the first site of the chain to the left lead, and the last site to the right lead, assuming a bilinear coupling of the form
\begin{equation}
\label{eq:hsb}
\hat{H}_{FL} + \hat{H}_{FR}=\sum_{\lambda}(t_{\lambda L}\hat{a}^{\dagger}_{1}\hat{d}_{\lambda L}+t_{\lambda R}\hat{a}^{\dagger}_{N}\hat{d}_{\lambda R}+ {\rm h.c}),
\end{equation} 
with $t_{\lambda L}$, $t_{\lambda R}$ describing respectively the amplitude for electrons to tunnel from left and right lead onto the wire. Each bath is described by a spectral function 
\begin{align}
\label{eq:J_bath}
\mathfrak{J}_{\nu}(E)=2\pi\sum_{\lambda}|t_{\lambda \nu}|^2\delta(E-E_{\lambda \nu}).
\end{align}
We make use of the wide-band limit (WBL) approximation, taking spectral functions that are identical and independent of energy: $\mathfrak{J}_{\nu}(E)=\gamma$, for $\nu = L,R$. The non-equilibrium steady-state electric $J_e$ and heat $J_q$ currents can be obtained via the Landauer-B{\"u}ttiker integrals
 \begin{align}
 \label{eq:e_current0}
 	J_{e}&= \frac{2e}{h} \int dE\tau_{L R}(E)[f_L(E)-f_{R}(E)],   \\
 	 \label{eq:q_current0}
 	J_{q}&=\frac{2}{h} \int dE(E-\mu_L)\tau_{L R}(E)\nonumber \\
    	& \hspace{3cm}\times [f_L(E)-f_{R}(E)],
 \end{align}
 where $\tau_{LR}(E)$ gives the transmission probability from left to right, the factor 2 indicates the spin-degeneracy and $f_{\nu}(E)=\{1 + \exp[(E - \mu_{\nu})/k_B T_{\nu}]\}^{-1}$ is the Fermi-Dirac distribution of bath $\nu = L,R$, with $h$ and $k_B$ the Planck and Boltzmann constants and $e$ the elementary charge. The above results give the electric and heat currents in the left lead. Similar expressions hold for electric and heat currents in the right lead. In general, the heat current in the left and in the right lead may differ. However, in this work, we are interested in the linear response regime where the chemical potential difference $\Delta \mu = \mu_L-\mu_R$ and the temperature difference $\Delta T = T_L-T_R $ are small compared to the average $T$ and $\mu$. In this case, the two heat currents can be considered to be approximately the same.
 
 \begin{figure}[t]
\centering
\subfloat[]{\includegraphics[width=0.5\columnwidth]{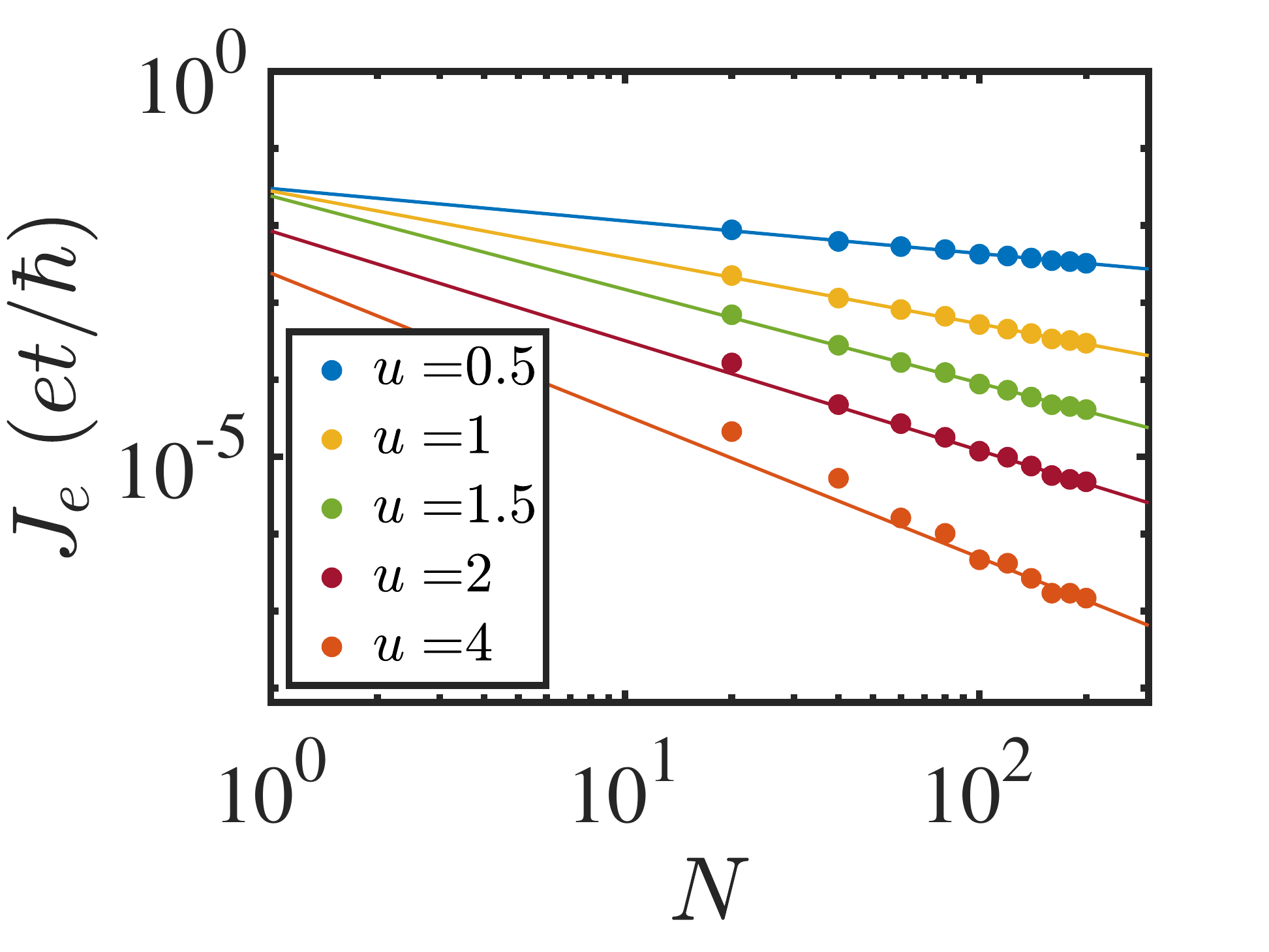}\label{fig:coje}}
\subfloat[]{\includegraphics[width=0.5\columnwidth]{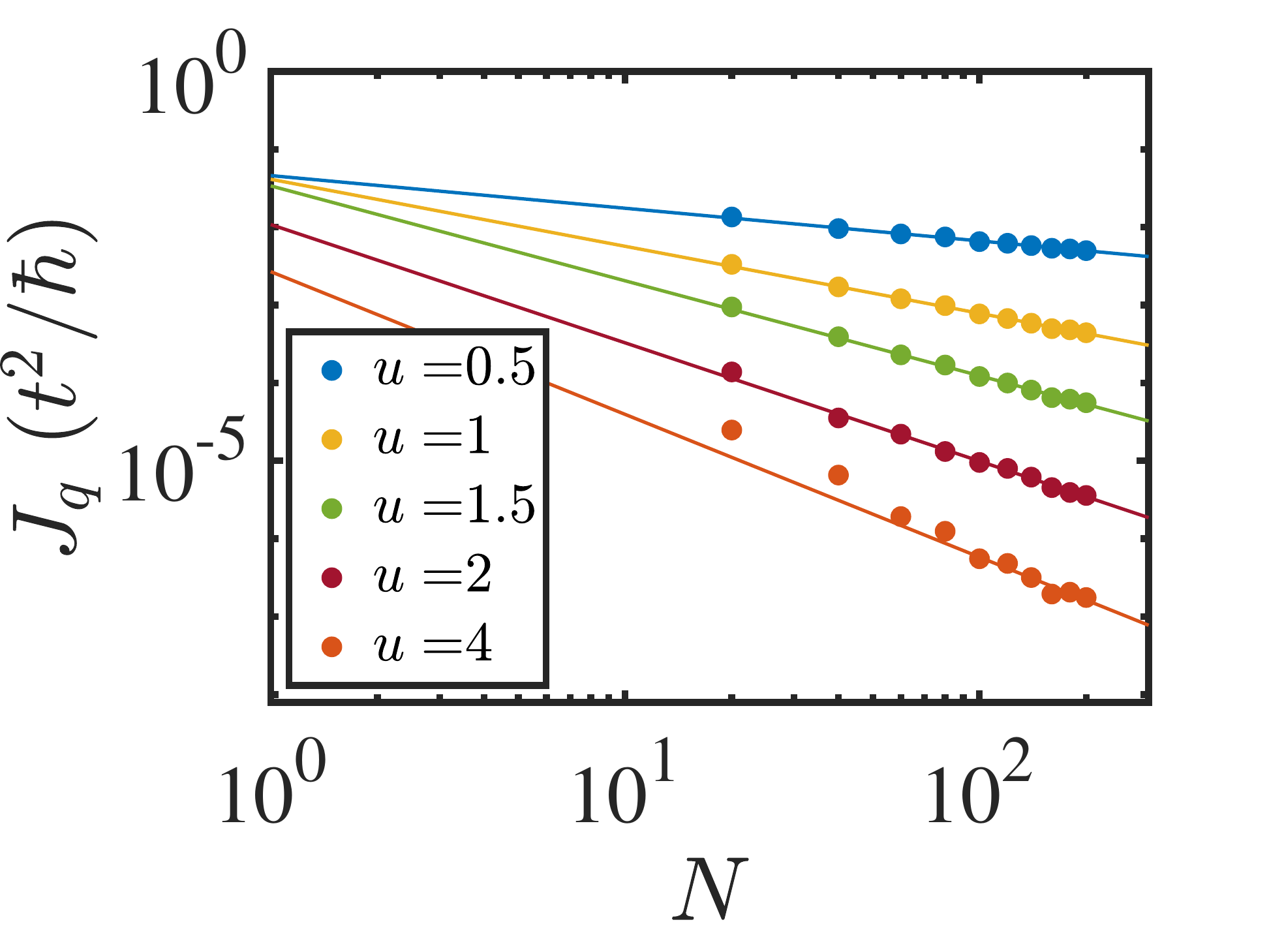}\label{fig:cojq}} \\
\subfloat[]{\includegraphics[width=0.5\columnwidth]{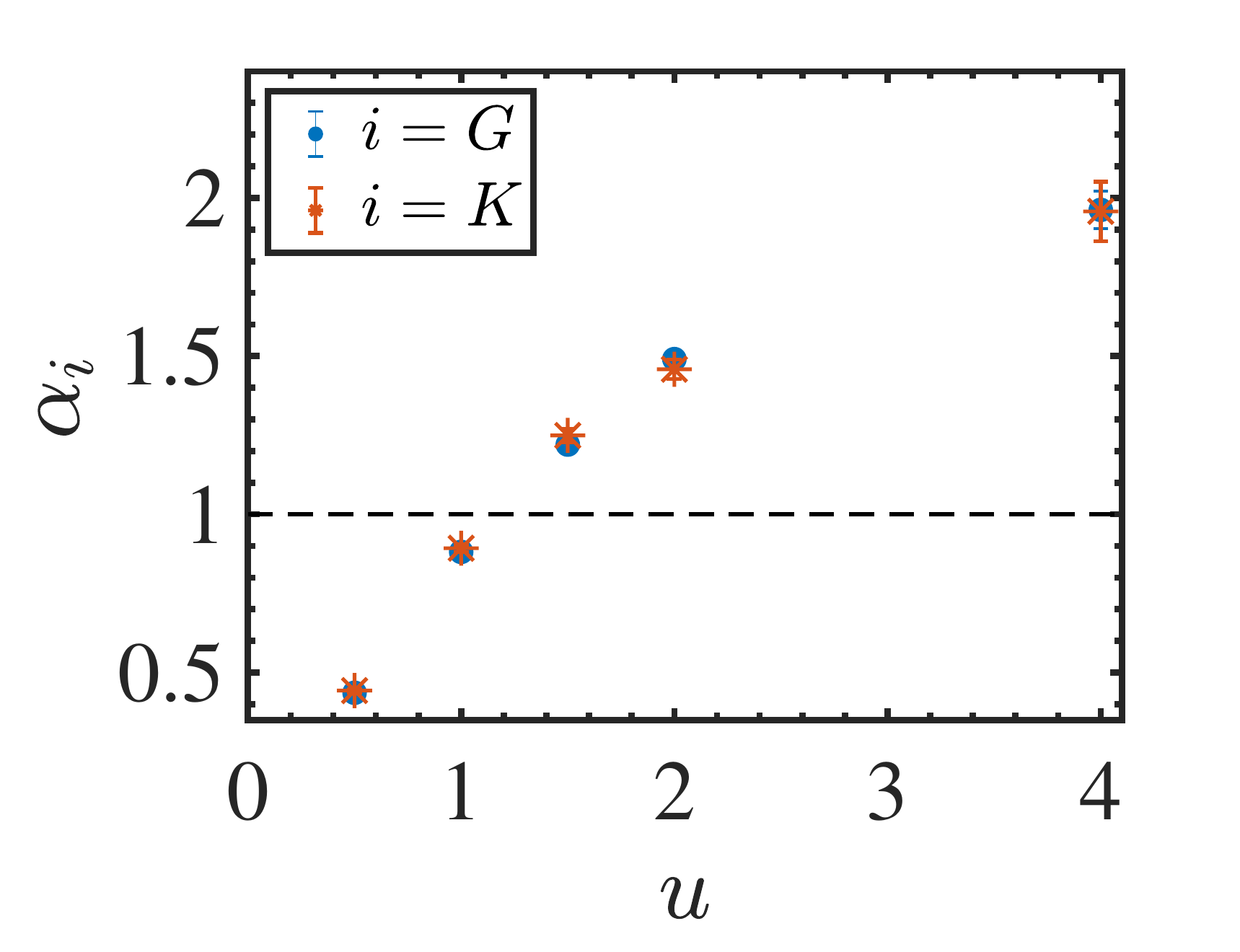}\label{fig:alphaGK}} 
\subfloat[]{\includegraphics[width=0.5\columnwidth]{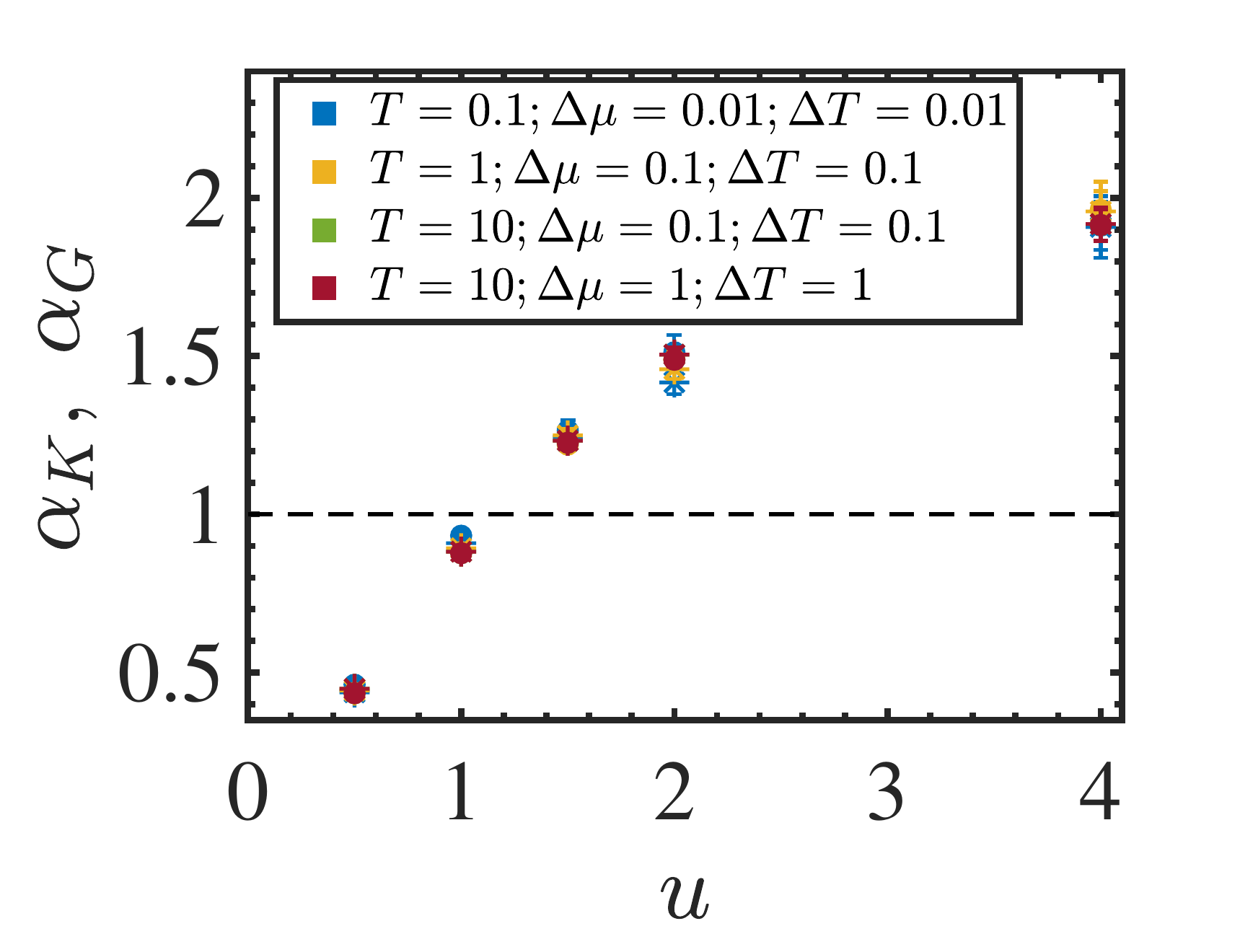}\label{fig:GKvari}}
\caption{(a)-(b) Scaling of coherent heat and electric currents in the Fibonacci model with hopping parameter $t=1$ and coupling to the baths $\gamma = 1$ at different potential strengths $u$, indicated in the color legend. The thermodynamic parameters are $T=1$, $\Delta T=0.1$, $\Delta \mu = 0.1$. The chemical potentials are respectively $\mu = -2, -2.4, -2.8, -3.3, -5.2$. (c) Scaling exponent extracted from the electric $G \sim N^{-\alpha_G}$ (blue dots) and thermal conductance $K \sim N^{-\alpha_K}$ (red stars) associated to the currents in (a)-(b), at the same parameters. The dashed line indicates the value of $\alpha$ at which transport is diffusive. The error bars are given by the asymptotic error in the fits. (d) Scaling exponents for $G$ (dots) and $K$ (stars) computed in different thermodynamic configurations given by the colors in the legend. We notice that they do not depend on the thermodynamic configurations.}
\end{figure} 

 The details of the Hamiltonian are encoded in the transmission function $\tau_{L R}(E)$, describing the probability for a particle at energy $E$ to be transferred from reservoir $L$ to reservoir $R$ via the central system. We compute the transmission via the retarded single particle non-equilibrium Green's function (NEGF), which is defined as
 \begin{equation}
\label{eq:gf}
    \mathbf{G}(E) = \bigl[ E \hat{\mathds{1}} - \mathbf{H} - \sum_{\nu} \mathbf{\Sigma}_{\nu}(E)  \bigr]^{-1},
\end{equation}
where $\hat{\mathds{1}}$ is the $N\times N$ identity matrix, $\mathbf{H}$ is the $N\times N$ tridiagonal matrix defined by writing the system Hamiltonian as $\hat{H}_F=\sum_{n,m=1}^N \mathbf{H}_{nm}\hat{a}_n^\dagger \hat{a}_m$ and $\hat{\Sigma}_{\nu}(E)$ is the $N\times N$ self-energy matrix for the $\nu$th reservoir attached to the system. After introducing the level-width functions $\mathbf{\Gamma}_{\nu}(E)=i(\mathbf{\Sigma}_{\nu}^\dagger(E)- \mathbf{\Sigma}_{\nu}(E))$, the transmission function is computed from $\tau_{LR}(E) = \text{Tr} \bigl\{ \mathbf{\Gamma}_{L}(E) \mathbf{G}^{\dagger}(E) \mathbf{\Gamma}_{R}(E) \mathbf{G}(E)  \bigr\}$~\cite{ryndyk2016, datta1997, Meir1992}. In WBL approximation, self-energies are independent of energy. For our set-up, their representation on the lattice basis has only one non-zero element each, given by $\left[\hat{\Sigma}_{L}(E)\right]_{11}=\left[\hat{\Sigma}_{R}(E)\right]_{NN} = - i\gamma/2$. In the following, we work in a regime of intermediate system-bath coupling, $\gamma=t=1$. However, the choice of $\gamma$ is largely immaterial. It has previously been shown that within linear response and WBL approximation, modifying $\gamma$ in this set-up rescales the currents without qualitatively affecting the transport behaviour\tcite{chiaracane}.

The Fibonacci model has a fractal spectrum, with critical single-particle eigenfunctions \tcite{jagannathan2021, mace2016, hiramoto1988, Zhong_1995}. This fractality of the spectrum is reflected on the transmission function $\tau_{LR}(E)$. An example of the transmission function for a chosen value of Fibonacci potential strength is shown in Fig.\tref{fig:transmission}. The calculation of currents requires an integration over energy of the transmission function multiplied by the Fermi-Dirac distributions. Due to the near-discontinuous nature of the transmission, this integration becomes challenging for large system sizes. Nevertheless, the system sizes we have been able to access are large enough to extract the asymptotic transport exponents $\alpha_G$ and $\alpha_K$. In Fig.\tref{fig:coje} and Fig.\ref{fig:cojq}, we show $J_e$ and $J_q$ as function of system size $N$ at different Fibonacci potential strengths $u=0.5, 1.0, 1.5, 2.0, 4.0$. The thermodynamic parameters are $T=1.0$, $\Delta \mu = 0.1$ and $\Delta T = 0.1$. We select different chemical potentials for every value of $u$, since the choice of $\mu$ along the energy axis affects only a pre-factor in the currents and not their scaling exponent, leaving the plots qualitatively equivalent. We observe in Fig.\tref{fig:alphaGK} that the transport exponents $\alpha_G$ (blue dots) and $\alpha_K$ (red stars) collapse onto the same trend. In Fig.\tref{fig:GKvari}, we show that this data collapse occurs independently of temperature. The Landauer-B{\"u}ttiker formalism  has not only the advantage to allow the study of charge and heat currents at finite temperature but in what follows it will also allow us to study the effect of dephasing in a systematic way by introducing B{\"u}ttiker probes~\tcite{buttiker1986}.
\section{Dephasing}
\label{sec:deph}

\subsection{B{\"u}ttiker probes}
The idea of introducing additional electron reservoirs as probes to mimic dephasing noise was first described by B{\"u}ttiker\tcite{buttiker1986}, and then applied to extended conductors by D'Amato and Pastawski\tcite{cattena2010, dapasta}. The additional reservoirs are treated as conventional baths, which receive particles and re-introduce them into the central system after scrambling their phase. The probes are \ov fictitious'' in the sense that their particle distributions are self-consistently determined in such a way to mimic different types of incoherent scattering processes, depending on the conditions implemented on the currents. Incoherent elastic scattering, where the electrons lose memory of their phase but conserve their energy, is recreated by cancelling the contribution to the electric current at each energy with the so-called \ov dephasing probe''\tcite{kilgour2015B}. When a chemical potential bias is applied to the system, incoherent inelastic scattering is introduced by setting the net electric currents towards each probe to zero, using a \ov voltage probe''\tcite{kilgour2015, Kim2016, kim2017}. If a temperature bias is also present, we encode non-dissipative inelastic scattering by further cancelling the heat currents going in to the probes. In this case, the average transfer of charge and heat from and towards the probes is zero, but single electrons exchange energy and momentum besides losing phase coherence. In this work, we implement this so-called \ov voltage-temperature probe''\tcite{korol2016}. It should be noted that this is fundamentally different from local pure dephasing Lindblad dissipators, most often used in the context of quantum information, which allow energy transfer even on average.

The configuration to study incoherent transport can be then described again by \teqref{eq:H}. However, the index $\nu$ covers now both \ov real'' left ($L$) and right ($R$) baths, and the $N$ probes, $\nu=L, R, 1, ..., N$. Each probe $n=1,...,N$ is a fermionic bath with Hamiltonian analogous to \teqref{eq:H_bath} and spectral function analogous to \teqref{eq:J_bath}, and is coupled to the $n$-th site of the chain through
\begin{equation}
\label{eq:hsp}
\hat{H}_{Fn}= \sum_{\lambda} ( t_{\lambda n}\hat{a}^{\dagger}_{n}\hat{d}_{\lambda n}+ {\rm h.c}),
\end{equation}
with $t_{\lambda n}$ the amplitude for electrons to tunnel from the $n$-th lead onto the wire.
The electric and heat currents flowing through the system are given by the Landauer-B{\"u}ttiker integrals in \teqsref{eq:e_current0}{eq:q_current0} extended to multiple terminals
 \begin{align}
 \label{eq:e_current}
 	J_{e}&= \frac{2e}{h} \sum_{\nu} \int dE\tau_{L \nu}(E)[f_L(E)-f_{\nu}(E)],   \\
 	 \label{eq:q_current}
 	J_{q}&=\frac{2}{h}\sum_{\nu} \int dE(E-\mu_L)\tau_{L \nu}(E)\nonumber \\
    	& \hspace{3cm}\times [f_L(E)-f_{\nu}(E)].
 \end{align}
The collection of transmission functions $\tau_{\nu'\nu}(E)$ is found via a generalization of the NEGF approach to a multi-terminal set-up. 
They are given by the following generalized formula~\cite{ryndyk2016, datta1997, Meir1992}
\begin{equation}
\tau_{\nu \nu'}(E) = \text{Tr} \bigl\{ \mathbf{\Gamma}_{\nu}(E) \mathbf{G}^{\dagger}(E) \mathbf{\Gamma}_{\nu'}(E) \mathbf{G}(E)  \bigr\},
\end{equation}
where the retarded single particle non-equilibrium Green's function (NEGF)  $\mathbf{G}(E)$ was defined in \teqref{eq:gf}. The indices $\nu, \nu'=L, R, 1, ..., N$ here run over the real left ($L$) and right ($R$) baths, and the B{\"u}ttiker probes ($1, ..., N$). The self-energies of \teqref{eq:gf} associated with the probes ($\nu = 1,...N$) are given in WBL approximation by one constant non-zero element matrices when in lattice basis, $\left[\hat{\Sigma}_{n}(E)\right]_{nn}=- i\gamma_d/2$. In the following, we will refer to the system-probe coupling parameter $\gamma_d$ as \ov dephasing strength''. However, as mentioned at the beginning of this section, the conditions implemented on the currents mimic incoherent inelastic scattering events, leading to energy relaxation (at single electron level, but not on average) beside the loss of phase coherence. Given the WBL approximation and the structure of the bilinear coupling with the central system in \teqref{eq:hsp} and \teqref{eq:hsb}, the generalized transmission functions  can be simplified as
\begin{align}
\label{eq:taus}
    \tau_{LR}(E) &= \gamma^2 \ | [\mathbf{G}(E) ]_{1N}|^2 \\
\tau_{nL}(E) &=  \gamma \gamma_d \ |[\mathbf{G}(E)]_{n1}|^2  \\
        \tau_{nR}(E) &=  \gamma \gamma_d \ \ |[\mathbf{G}(E)]_{nN}|^2 \\
\tau_{nn'}(E) &=  \gamma^2_d \ \ |[\mathbf{G}(E)]_{nn'}|^2.
\end{align}
We assign $\tau_{\nu \nu}(E)=0$, since these terms do not contribute to the currents, and $\tau_{\nu \nu'}(E)=\tau_{\nu' \nu}(E)$, since the tunnelling process is symmetric.
 However, the number of transmissions to compute at every energy $E$ grows as $N^2$, limiting our study to $N \sim 200$. Despite this we find that our numerics are well converged at this system size and allow for an accurate extraction of transport exponents.

The only formal difference between real baths and the probes is that temperature $T_n$ and chemical potential $\mu_n$ of the latter are not free parameters, but self-consistently determined by imposing charge conservation and absence of heat dissipation on each probe, as follows
\begin{align}
\label{eq:probe_cond}
    	J_{e, n}&= \frac{2e}{h} \sum_{\nu} \int dE\tau_{n \nu}(E)[f_n(E)-f_{\nu}(E)] = 0, \\
    	\label{eq:probe_cond2}
    	J_{q, n}&=\frac{2}{h}\sum_{\nu} \int dE(E-\mu_{n})\tau_{n \nu}(E) \nonumber \\
    	& \hspace{3cm}\times [f_n(E)-f_{\nu}(E)] = 0.
\end{align}
However, these 2$N$ non-linear equations do not posses a proof of existence and uniqueness of the solution, contrary to the case of the voltage probe\tcite{jacquet2012, kilgour2016B}. We restrict then the study to linear response regime, as suggested by the algorithm in Ref.\tcite{probezt_rev}, which instead gives 2$N$ linear equations that can be solved relatively easily. To extract the transport coefficients defined in \teqsref{eq:currcoff}{eq:currcoff2} we first calculate electric and heat currents setting $\Delta T=0$, $\Delta \mu \neq 0$, and then calculate the same, setting $\Delta T\neq 0$, $\Delta \mu = 0$. From \teqsref{eq:currcoff}{eq:currcoff2}, we see that the first calculation allows extraction of $G$ and $\Pi$. Knowing $G$ and $\Pi$, the second calculation allows extraction of $S$ and $K$. For each of these calculations,  we solve the corresponding linear system of equations and plug the set of solutions $\{T_n, \mu_n \}$ into \teqsref{eq:e_current}{eq:q_current}. In the next subsection, we investigate the possibility of dephasing-enhanced transport in the Fibonacci model in presence of the Buttiker probes.

\subsection{Dephasing-enhanced transport}
\label{sec:deph_transp}

\begin{figure}[t]
\centering
\subfloat[]{\includegraphics[width=0.5\columnwidth]{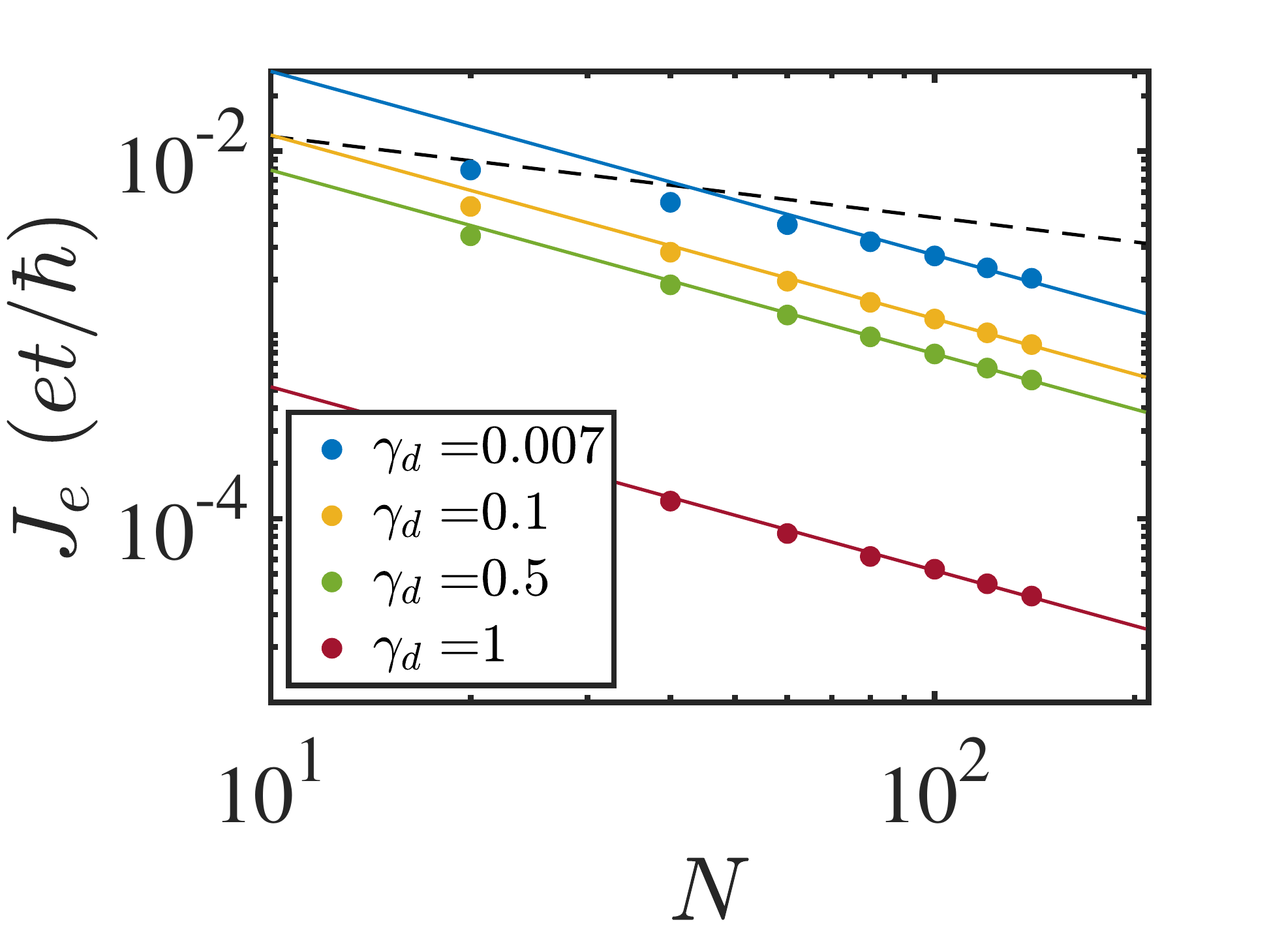}\label{fig:je_u05}}
\subfloat[]{\includegraphics[width=0.5\columnwidth]{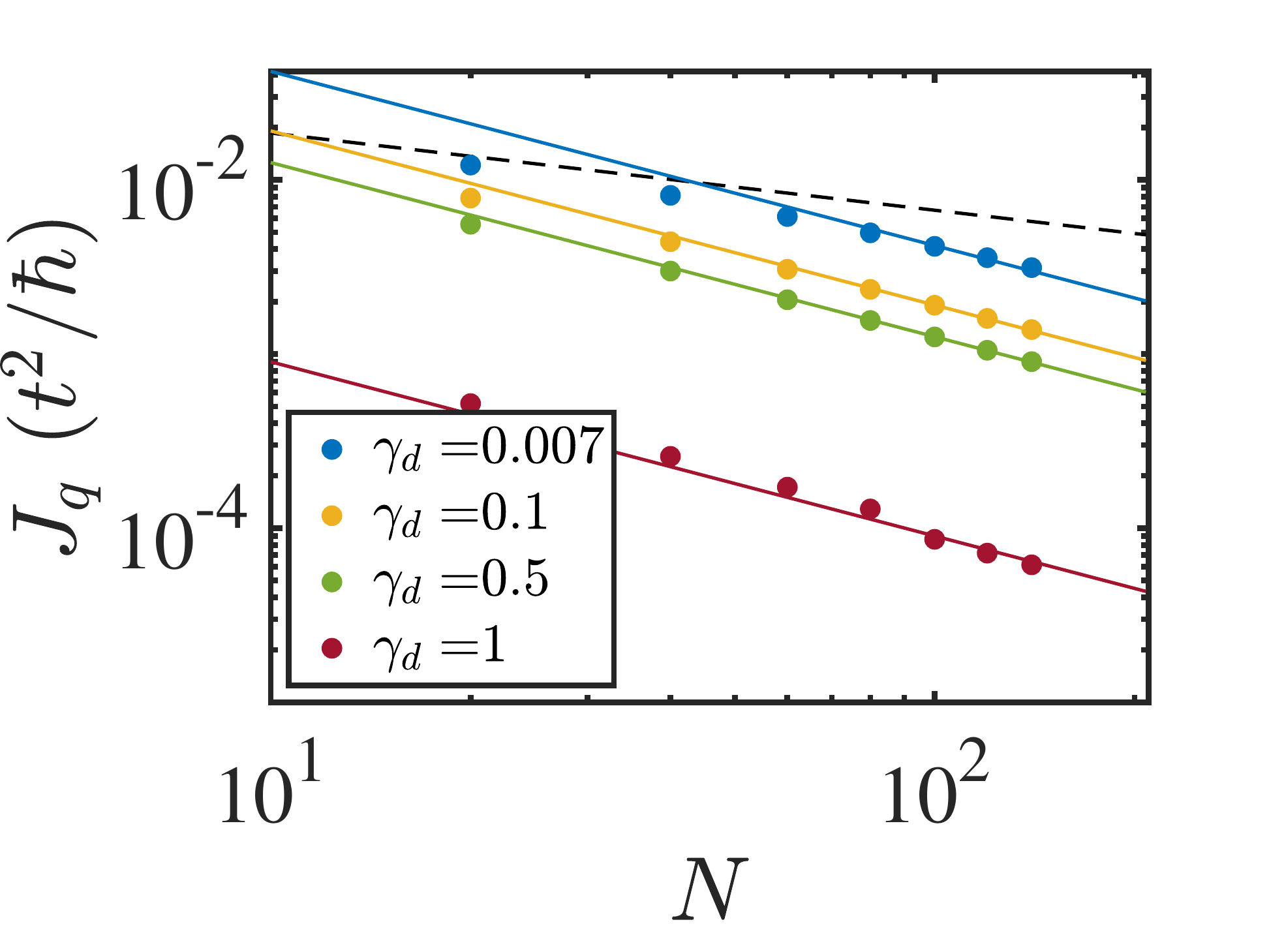}\label{fig:jq_u05}} \\
\subfloat[]{\includegraphics[width=0.5\columnwidth]{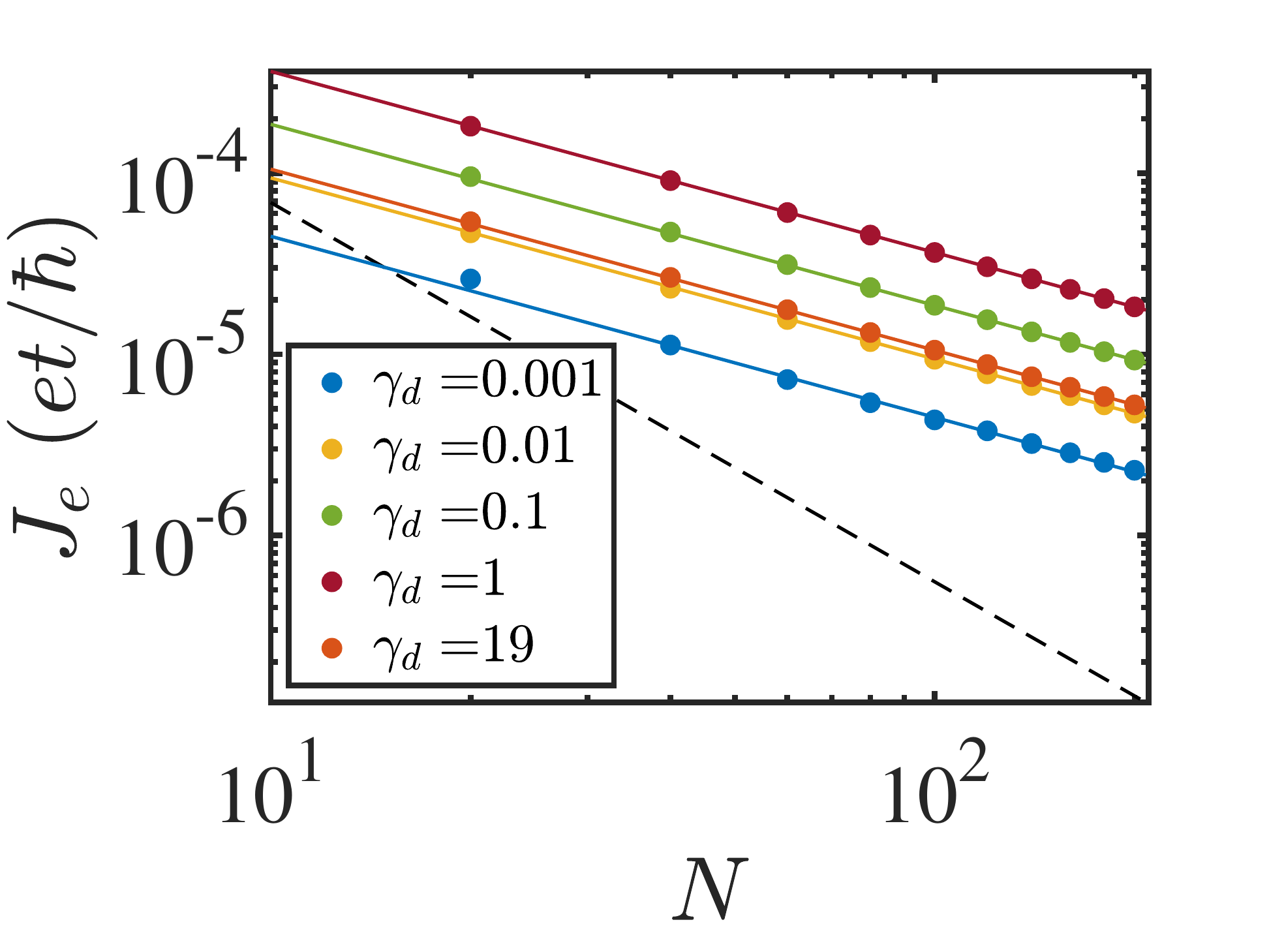}\label{fig:je_u4}} 
\subfloat[]{\includegraphics[width=0.5\columnwidth]{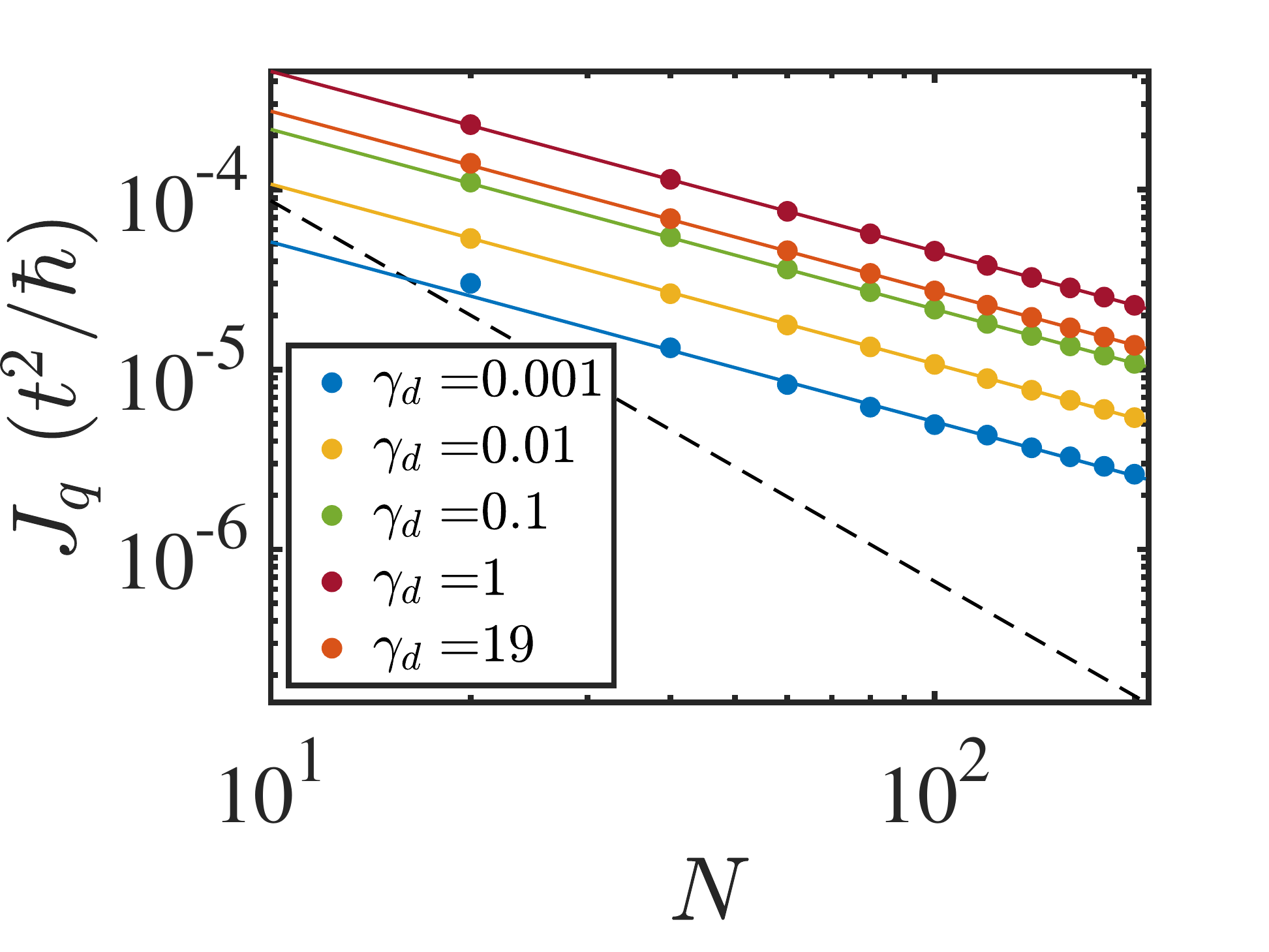}\label{fig:jq_u4}}
\caption{Electric (a)-(c) and heat (b)-(d) currents in Fibonacci chains of length $N$ at various dephasing strengths $\gamma_d$, indicated in the legends. The dashed line shows the corresponding currents at zero dephasing. Currents become diffusive at any $\gamma_d \neq 0$, so that transport slows down in the superdiffusive regime for $u = 0.5$ (top panel), while is enhanced in the subdiffusive regime for $u=4.0$ (bottom panel). The thermodynamic parameters are $T=1.0$, $\Delta T =0.1$, and $\Delta \mu =0.1$.}
\label{fig:enhancedcurr}
\end{figure} 

\begin{figure*}[t]
\centering
\subfloat[]{\includegraphics[width=0.62\columnwidth]{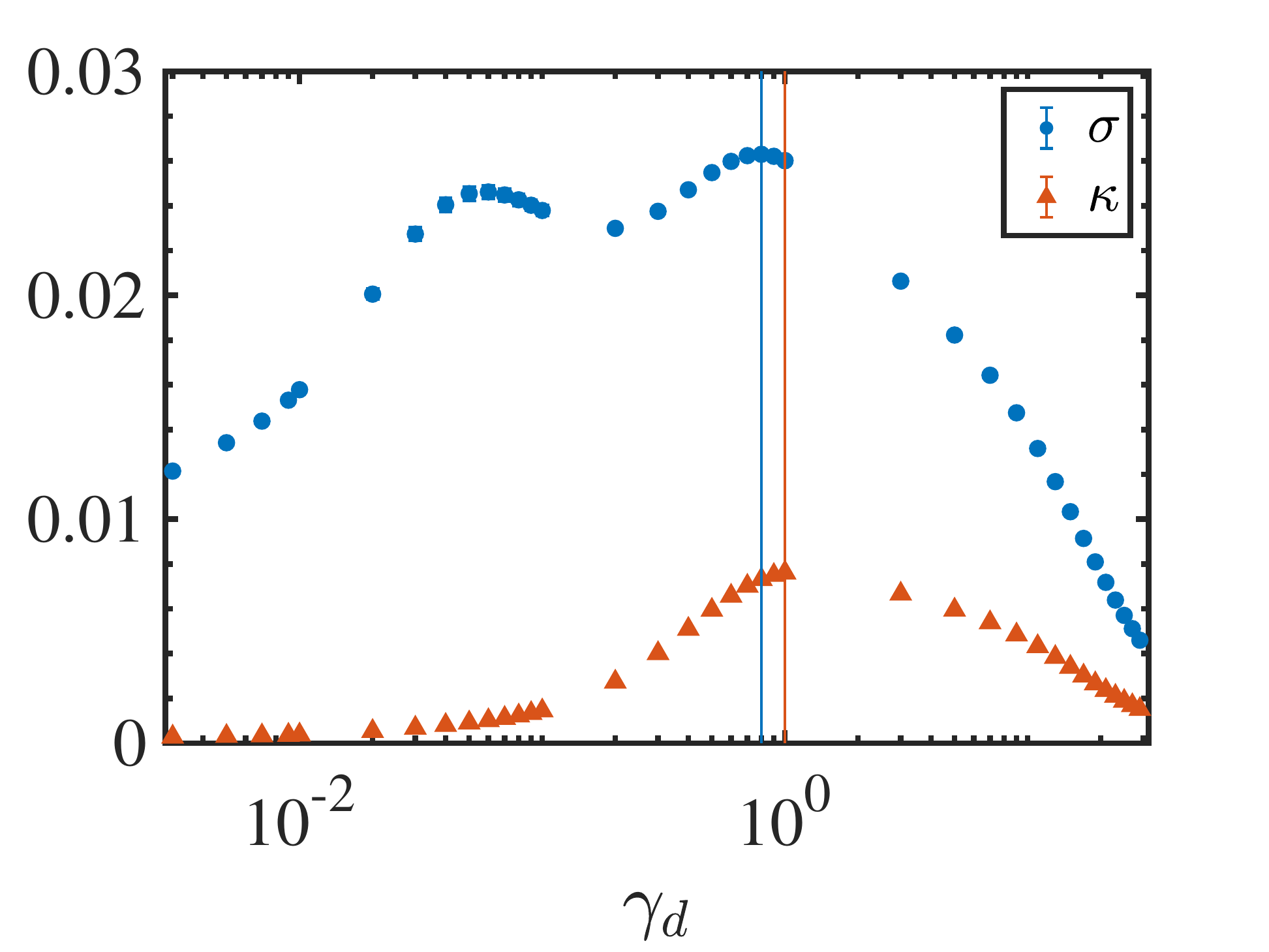}\label{fig:tc1}}
\subfloat[]{\includegraphics[width=0.62\columnwidth]{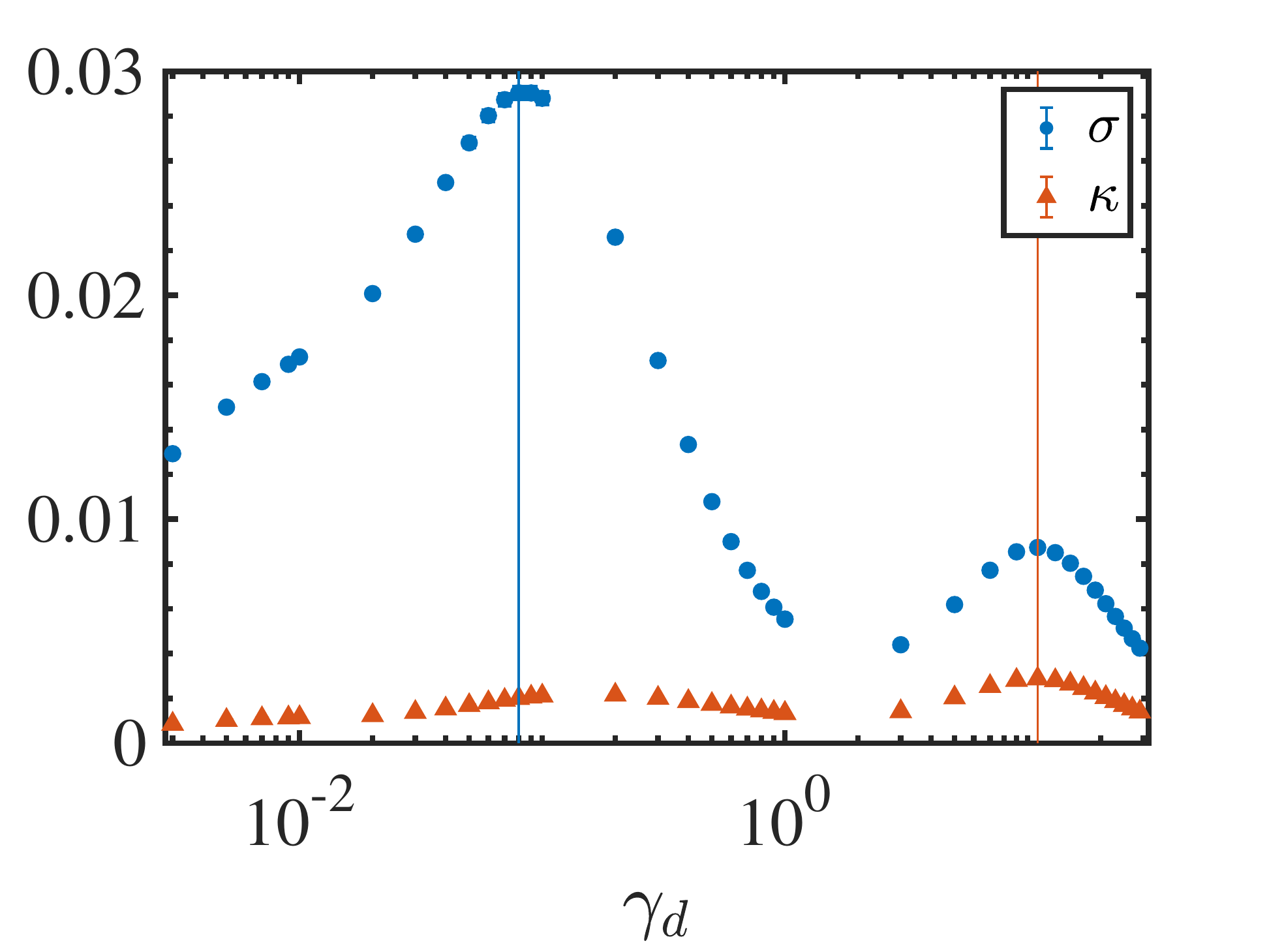}\label{fig:tc2}}
\subfloat[]{\includegraphics[width=0.62\columnwidth]{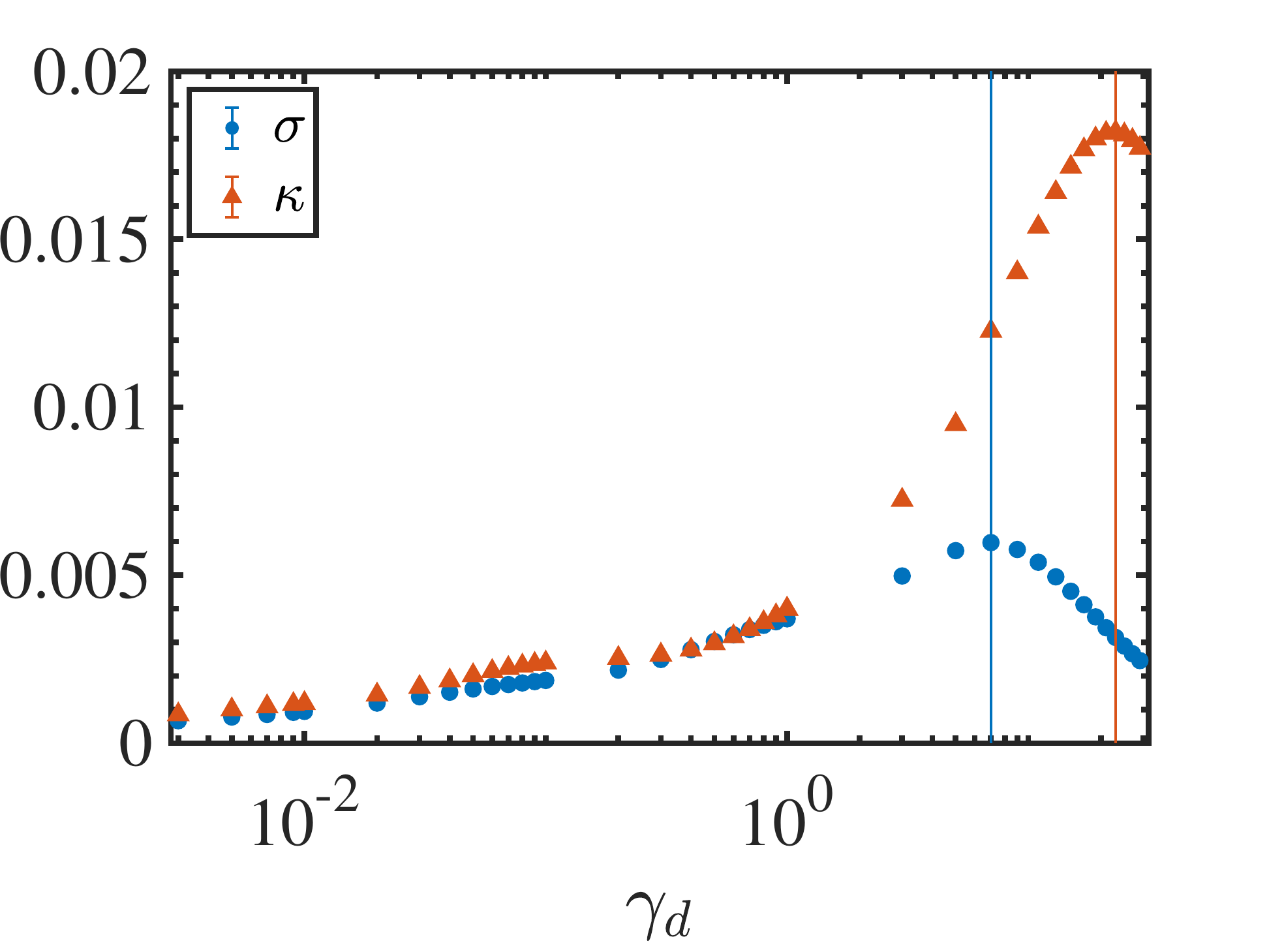}\label{fig:tc3}} 
\caption{The electric (blue) and thermal (red) conductivities extracted from the scaling of the conductances up to a length of $N=200$, with $u=4.0$. The continuous lines highlight the dephasing strength $\gamma_d$ that maximises the corresponding conductivity. The plots are at different thermodynamic configurations: in (a)-(b), $T=0.1$ and $\mu$ is taken at two different points in the energy spectrum, respectevely $\mu=-5.2$ and $\mu=4.3$, while in (c), $T=10$ and the choice of $\mu$ becomes irrelevant (for the specific plot we show $\mu = -5.2$). The error bars on each data point, given from the asymptotic error in the linear fit, are smaller than dot size and not visible in the plots.}
\label{fig:conductivities}
\end{figure*}

 A heuristic argument to understand the behaviour of the infinite-temperature conductivity after adding dephasing was introduced in Ref.\tcite{znidaric2017} for spin transport with dephasing and dissipation modelled via Lindblad equations. Here, we revisit the argument considering electric current under a voltage bias. The electric current induced by the voltage bias $\Delta \mu$ is defined as $J_e= \sigma (N) \Delta \mu/ N \sim N^{-\alpha_G}$, where $\alpha_G$ is the transport exponent related to the conductance $G$ in the absence of dephasing. It is known that sufficient dephasing changes anomalous transport behavior to normal diffusive behavior. For a given dephasing strength $\gamma_d$, one can associate a characteristic length $N_d$, beyond which coherence is quickly destroyed, so that the transport becomes diffusive with well-defined $\sigma(\gamma_d)$. This argument gives  
\begin{equation}
    \sigma(N, \gamma_d) \sim 
    \begin{cases}
    N^{1-\alpha_G} \ \ \ \ N < N_d \\
    \sigma(\gamma_d) \ \ \ \ N > N_d
    \end{cases},
\end{equation}
The behaviour should be continuous across $N_d$,
so that at $N=N_d$ it must hold that $\sigma(\gamma_d) \sim N_d^{1-\alpha_G}$. Considering $\tau_d \sim 1/\gamma_d$ to be the time between incoherent scattering events, $N_d$ can be heuristically estimated by the spatial spread of a small perturbation in the system within this time in absence of coupling to baths \tcite{znidaric2017}. This gives $N_d \sim \gamma_d^{-1/(\alpha_G+1)}$. As a result, for small dephasing strength, we get the following dependence of conductivity on the dephasing strength,
\begin{equation}
    \sigma(\gamma_d) \sim N_d^{1-\alpha_G} \sim \gamma_d^{(\alpha_G -1)/(\alpha_G + 1)}.
\end{equation}
Thus, the dependence of conductance on the dephasing strength is dictated by the nature of transport in the absence of dephasing. If the transport in the absence of dephasing is either ballistic ($\alpha_G = 0$) or superdiffusive ($\alpha_G <1$), in the regime of small $\gamma_d$, the conductivity decays to zero as $\gamma_d$ increases. But in the case of subdiffusion ($\alpha_G>1$), the conductivity increases  and consequently reaches a maximum at intermediate $\gamma_d$ before decaying for large $\gamma_d$. Thus, dephasing enhanced transport is expected in the regime where the transport was subdiffusive in absence of dephasing.

Behavior consistent with above heuristic description has already been observed in various systems within the framework of local Lindblad equations, which can be thought to model the infinite temperature limit, and local pure dephasing Lindblad dissipators\tcite{znidaric2017, znidaric2013, mendoza2013,lacerda2021, znidaric2021more}. This includes a recent study on the Fibonacci model \tcite{lacerda2021}. We stress again that our set-up is fundamentally different from this class of descriptions. In the set-up of these previous works, energy exchange with the sources of dephasing is allowed, even on average. However, in our set-up with the voltage-temperature B{\"u}ttiker probes, both electric and heat currents into the probes are zero on average. Therefore, neither particle exchange nor energy exchange with the sources of dephasing are allowed on average. Despite this, we expect the heuristic phenomenology of dephasing enhanced transport given above to hold in our set-up. Moreover, although the above phenomenology has been discussed in terms of electric conductivity,  we expect to see enhancement of thermal conductivity also as a function of dephasing strength, before it eventually decays to zero for large dephasing strength.

We now numerically explore the possibility of dephasing enhanced transport in the our set-up. To this end,
in Figs.\tref{fig:enhancedcurr}, we show the diffusive scaling of electric $J_e$ (left panels) and heat $J_q$ (right panels) currents at different $\gamma_d$ for potential strength $u=0.5$ (top panels) and $u=4.0$ (bottom panels). In the same figures, the dashed line indicates the value of currents in the coherent case. We verify, as evident in the bottom panels, that dephasing enhances heat and electric transport at the potential strength which would otherwise determine subdiffusion, $u=4.0$. The plots are realized for specific $\mu$, $T=1.0$, and $\Delta \mu = \Delta T = 0.1$, but changing the thermodynamic variables of the leads does not alter the results in any qualitative way. 

Next, we look at the electric and thermal conductivities.
We extract the conductivities $\sigma$ and $\kappa$ from the linear fits of respectively $\log{G}$ and $\log{K}$ versus $-\log{N}$ up to $N=200$, for different values of $\gamma_d$. While scanning the thermodynamic parameter space, we notice a remarkably sensitive behaviour of the conductivities to temperature $T$ and chemical potential $\mu$, which is more evident as we increase the potential strength $u$ in the subdiffusive regime. In Fig.\tref{fig:conductivities} we show $\sigma$ (in blue) and $\kappa$ (in red) as a function of $\gamma_d$ for $u=4.0$ at different choices of $T$ and $\mu$. In all plots, we see that both the electrical and the thermal conductivities initially increase with $\gamma_d$, while they go to zero for large $\gamma_d$, as expected from the heuristic argument above. We highlight the position of the highest values of $\sigma$ and $\kappa$ with continuous vertical lines of the same color. In Fig.\tref{fig:tc1} and Fig.\tref{fig:tc2} we set the temperature to $T=0.1$, and take two different values of $\mu$, respectively corresponding to the lower and top end of the spectrum.  Surprisingly, we observe the presence of multiple local maxima, whose heights and positions depend on the choice of $\mu$. The same kind of variety in the local peaks arises also at intermediate temperatures and for other choices of chemical potentials. On the other hand, at high temperatures, a single peak appears for each conductivity, with position and height independent of $\mu$, as shown in Fig.\tref{fig:tc3} for $T=10$. The presence of a single peak is in consistent with previous findings using Lindblad dephasing in Ref.\tcite{lacerda2021}. 

 Linear response transport properties of a fermionic system at chemical potential $\mu$ and temperature $T$ are usually governed by the spectrum of the system in the range of the energies $\mu \pm k_B T$, which is approximately the width of the derivative of the Fermi-Dirac distribution with respect to $\mu$. Thus, if $k_B T$ is much larger than the bandwidth of the system, transport coefficients become independent of $\mu$. This explains the observed $\mu$ independence of high temperature conductivities. On the other hand, this picture suggests that the presence of multiple $\mu$ dependent peaks at low temperatures is related to the structure of the effective spectrum given by the collection of transmission functions within the energy window $\mu \pm k_B T$. We therefore deduce that the fractal spectrum of the Fibonacci model, which gives the peculiar near-discontinuous transmission function in the coherent case (see Fig.\ref{fig:transmission}), is also the reason for the suprising multiple peaks in the conductivities as a function of $\gamma_d$. A more microscopic understanding, however, is difficult at finite temperatures. Instead, in the next subsection we discuss another surprising observation from the results, the violation of Wiedemann-Franz law.

\begin{figure}[t]
\centering
\subfloat[]{\includegraphics[width=0.5\columnwidth]{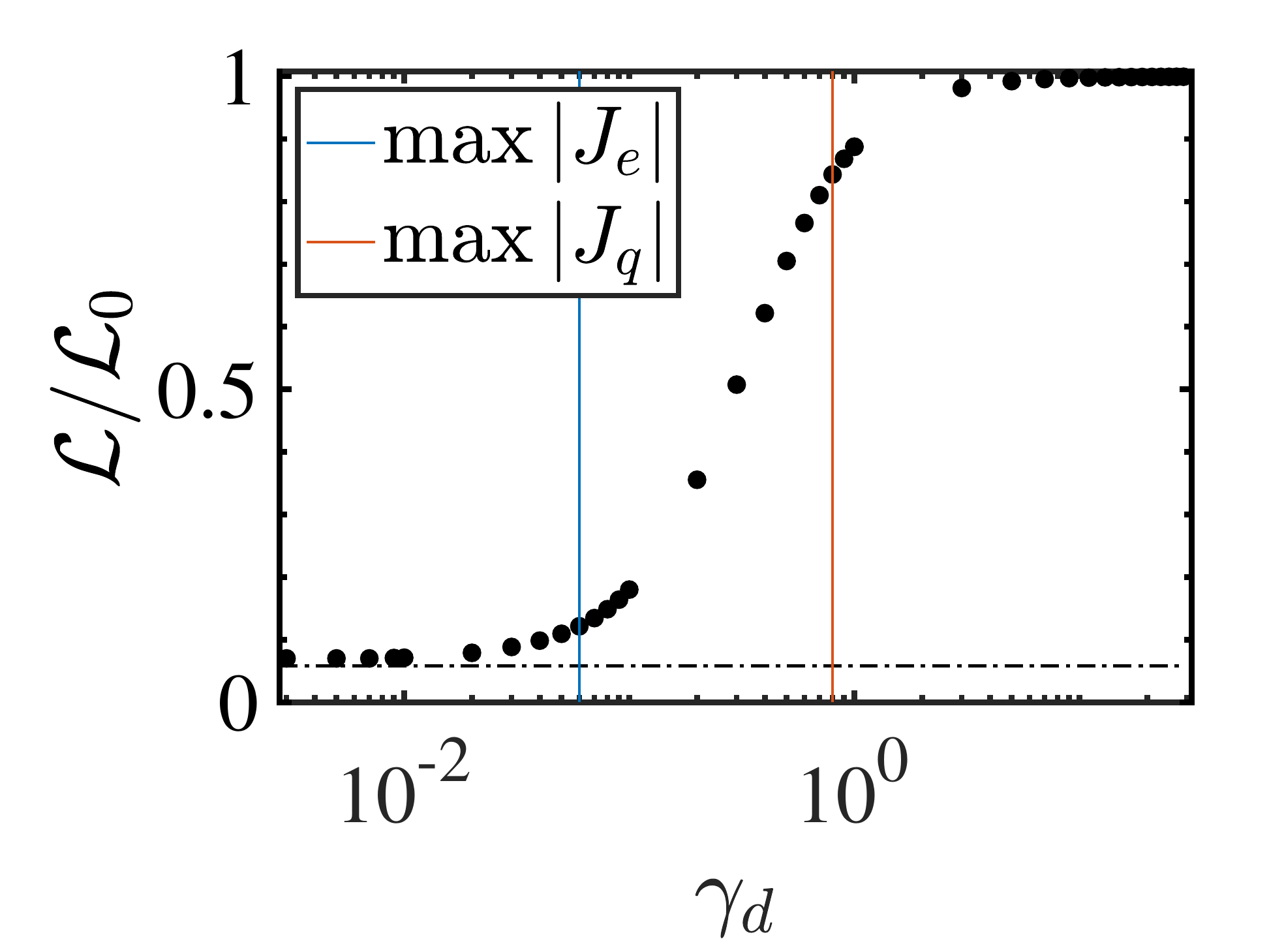}\label{fig:u4lor}}
\subfloat[]{\includegraphics[width=0.5\columnwidth]{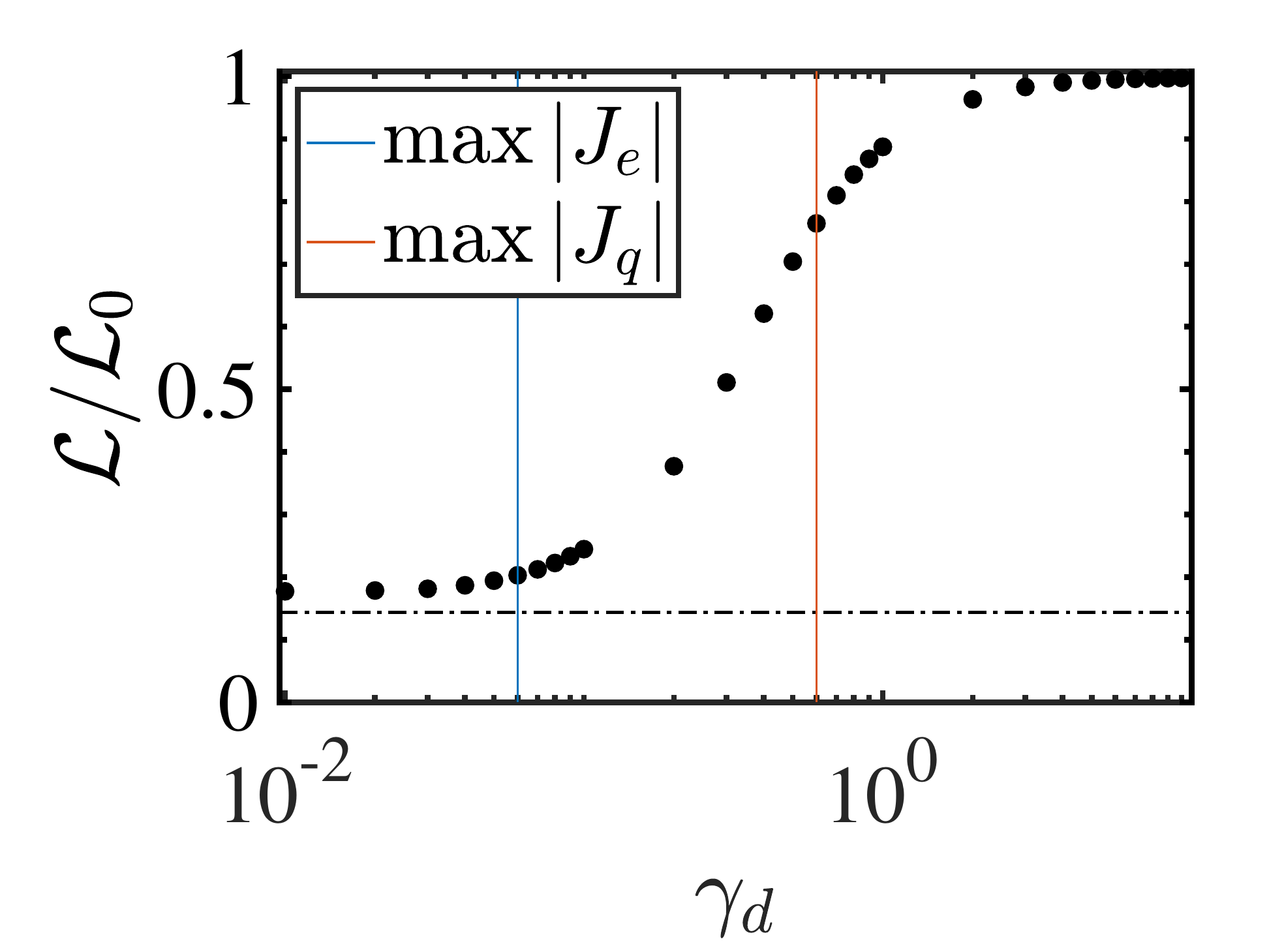}\label{fig:u2lor}} \\
\subfloat[]{\includegraphics[width=0.5\columnwidth]{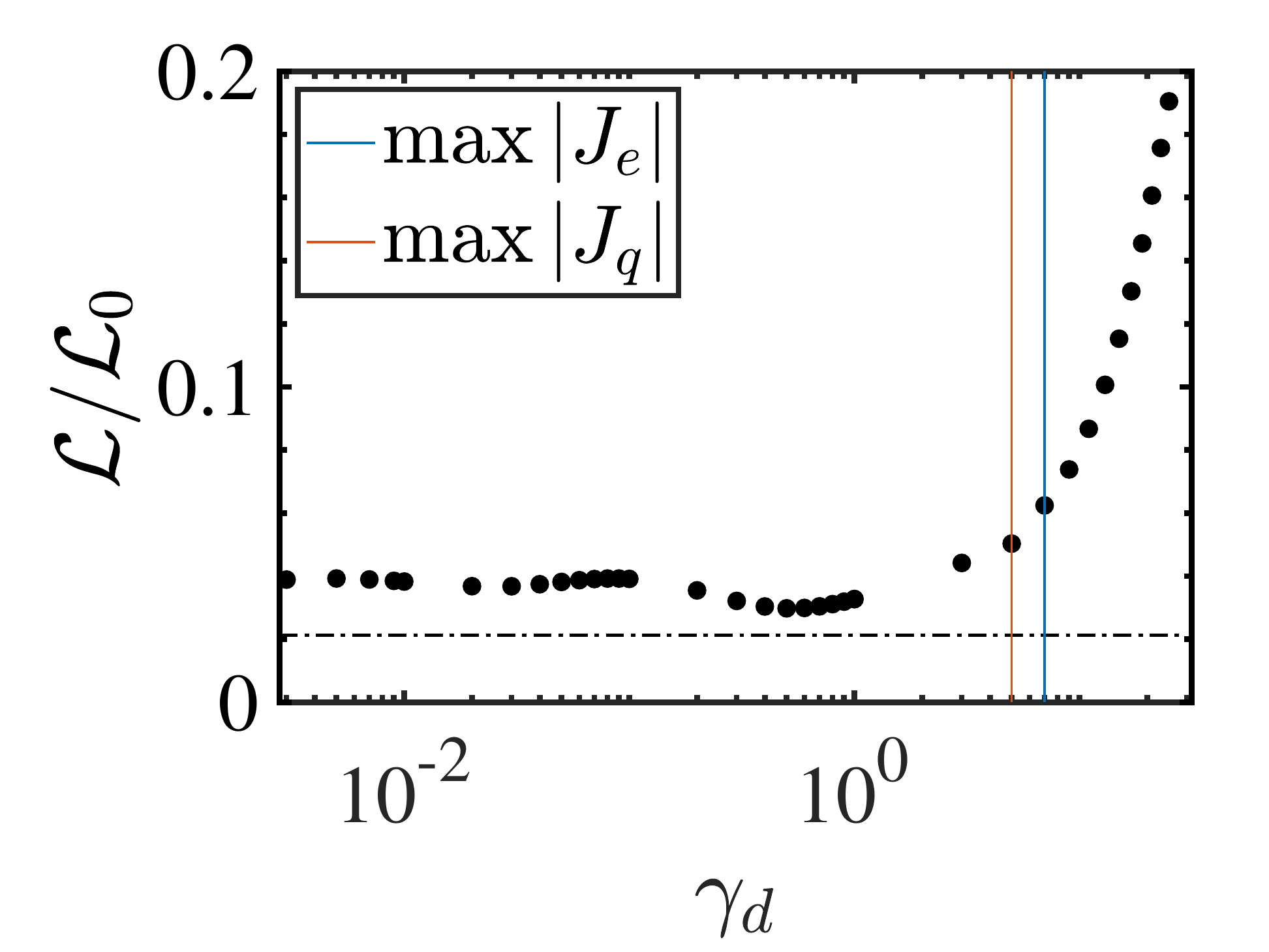}\label{fig:u4lor2}}
\subfloat[]{\includegraphics[width=0.5\columnwidth]{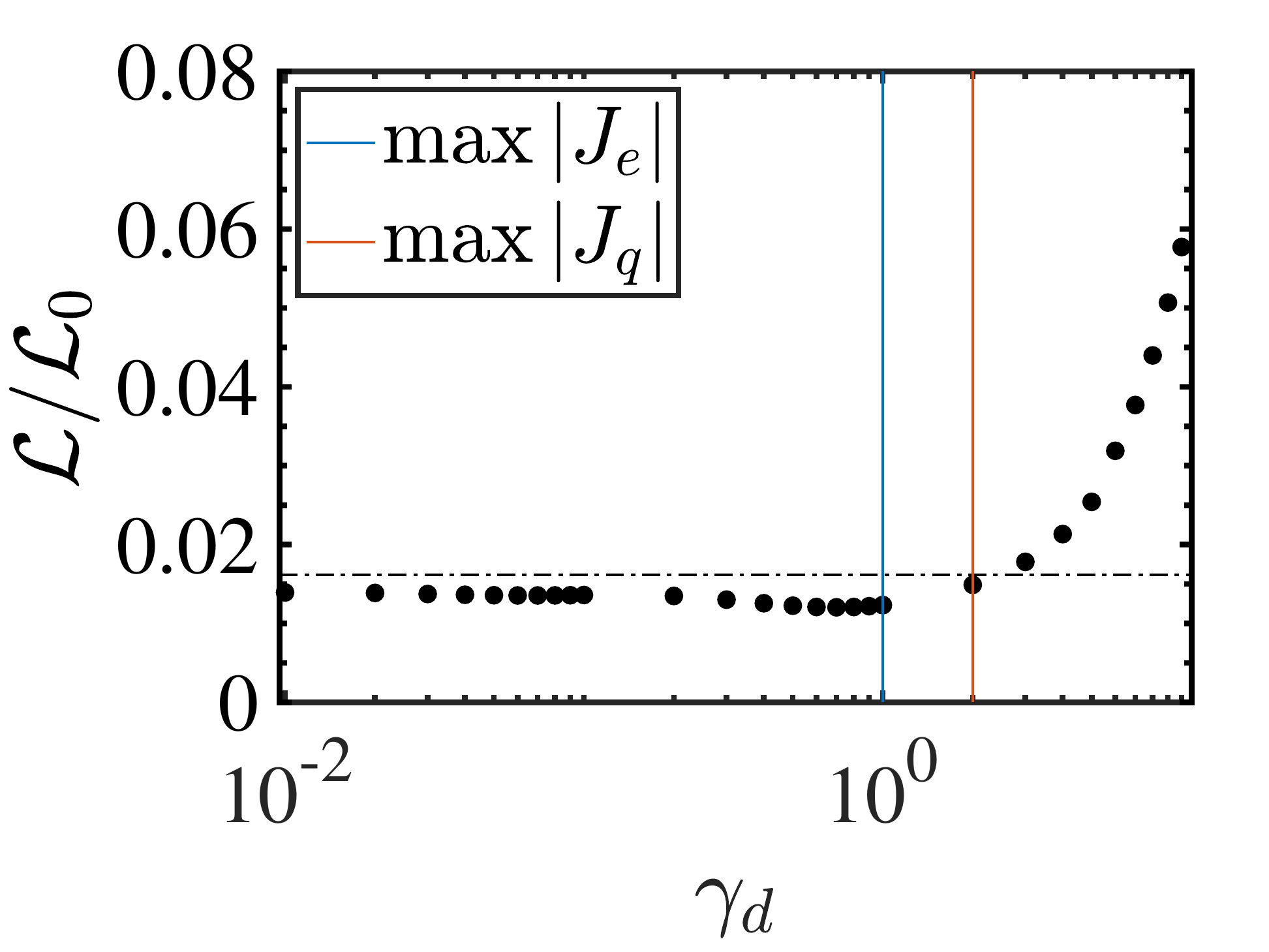}\label{fig:u2lor2}}
\caption{Ratio $\mathcal{L} = K/GT$ normalized to the Lorenz number $\mathcal{L}_0 = (\pi k_B)^2/3e^2$ for (a)~$u = 4.0$, $\mu = -5.2$, (b)~$u=2.0$, $\mu = -3.3$, at low temperature $T=0.1$, with $\Delta T = 0.01$, $\Delta \mu = -0.01$. The dashed line indicates the value at zero dephasing. The blue and red vertical continuous lines highlight respectively the position of the maxima of electric and heat current. In (c) and (d) we use the same parameters of the refrigerator configurations in (b)-(c) of Fig.\tref{fig:fridge}: (c)~$u=4.0, \mu = -5.2, T=10, \Delta T = 0.5, \Delta \mu = -1.0$, (d)~$u=4.0, \mu = -3.3, T=10, \Delta T = 1.0, \Delta \mu = -1.0$}.
\label{fig:lorenz}
\end{figure} 

\begin{figure}[t]
\centering
\subfloat[]{\includegraphics[width=0.5\columnwidth]{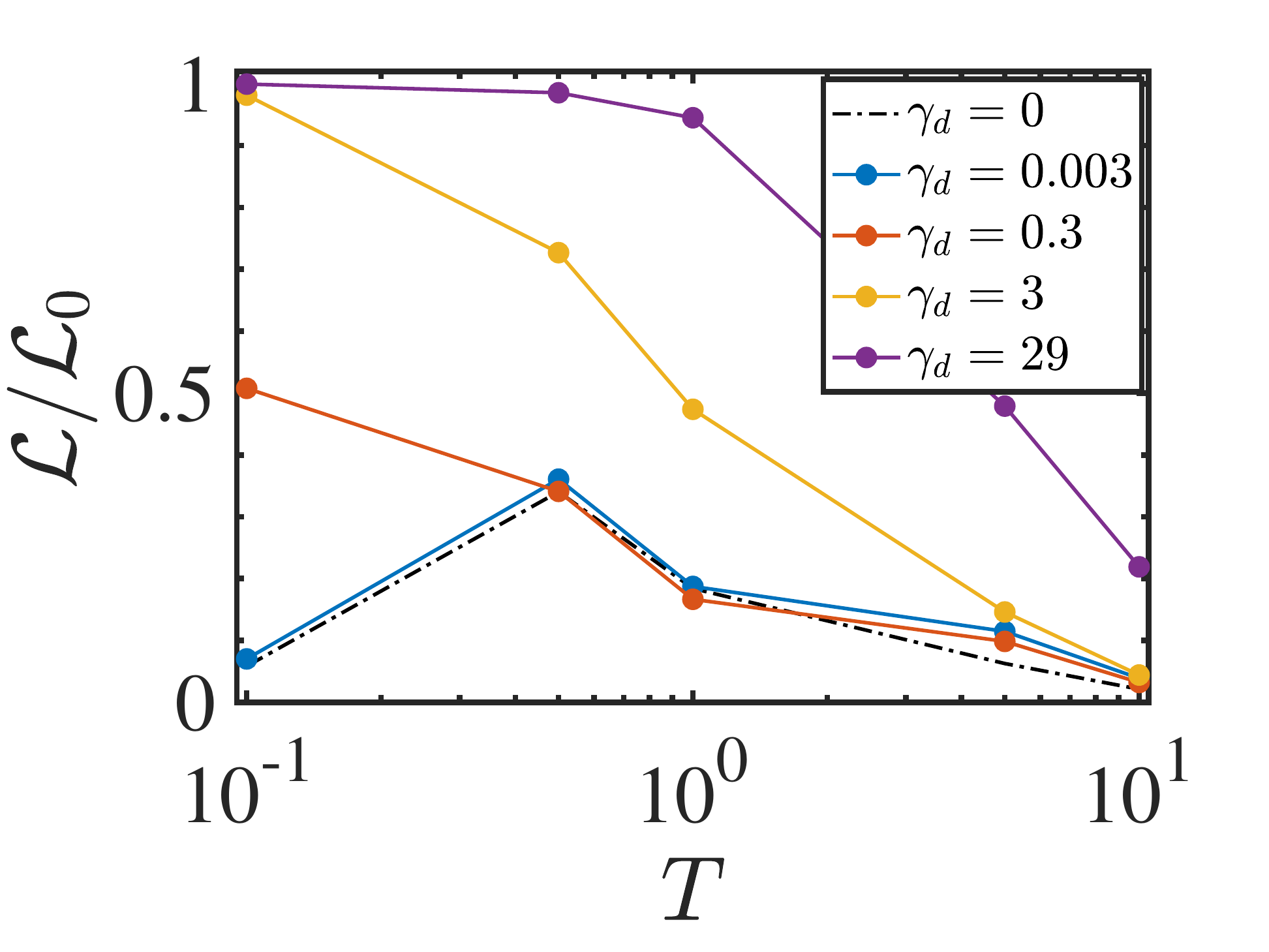}\label{fig:u4lor_vs_T}}
\subfloat[]{\includegraphics[width=0.5\columnwidth]{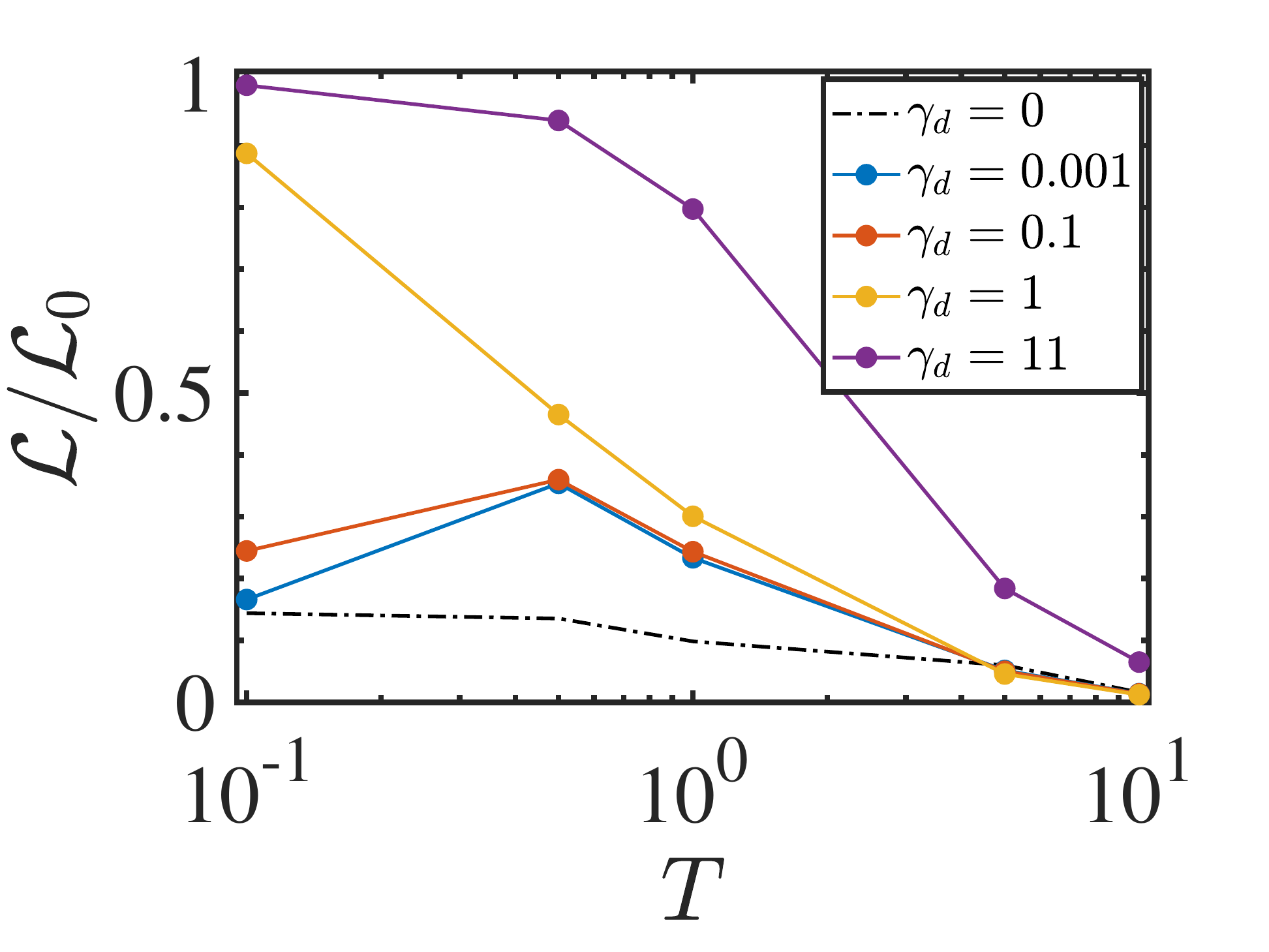}\label{fig:u2lor_vs_T}} 
\caption{Ratio $\mathcal{L}/\mathcal{L}_0$ for different choices of $\gamma_d$ as a function of temperature at (a) $u=2.0$, $\mu = -3.3$, (b) $u=4.0$, $\mu = -5.2$ with constant applied biases $\Delta \mu = 0.01$, $\Delta T = 0.01$.}.
\label{fig:lorenzT}
\end{figure}

\subsection{Violation of Wiedemann-Franz law}

The Wiedemann-Franz law states that in normal conductors at low temperatures the ratio of thermal conductivity over the product of electrical conductivity and temperature is a universal constant,
\begin{align}
    \frac{\kappa}{\sigma T}=\frac{K}{G T}=\mathcal{L}, ~~\mathcal{L}_0 = \frac{1}{3}\left(\frac{\pi k_B}{e}\right)^2.
\end{align}
The universal constant $\mathcal{L}_0$ is called the Lorenz number. This law shows that at a fixed temperature, electrical and thermal conductivities are proportional to each other. If transport is anomalous, this law need not hold, because the conductivities may not be well-defined in that case. Indeed, in the Fibonacci model in absence of dephasing, we find that the Wiedemann-Franz law, written in terms of the conductances, is violated. 

However, surprisingly, even in presence of dephasing, when the transport becomes diffusive and both the conductivities are well-defined, we see that Wiedemann-Franz law is still violated over a wide range. This remarkable fact is completely clear from Fig.\tref{fig:conductivities}, which shows that even at relatively low temperature $T=0.1$, the thermal and the electrical conductivities are not proportional to each other. In fact, we find that the maxima in the thermal and the electrical conductivities arise at different positions in parameter space, at both low and high temperatures. The violation of the Wiedemann-Franz law as a function of $\gamma_d$ at $T=0.1$ is explicitly shown in Fig.\tref{fig:u4lor} and in Fig.\tref{fig:u2lor}, respectively for $u=2.0$, and $u=4.0$. The $\mathcal{L}$ ratio is smaller than the Lorenz number for a wide range of $\gamma_d$, and it is restored to $\mathcal{L}_0$ only at $\gamma_d >> u$. At high temperatures, instead, as in Fig.\tref{fig:u4lor2} and in Fig.\tref{fig:u2lor2} the law is violated as expected for the entire range of $\gamma_d$ we have considered. We further analyze the deviation by visualizing $\mathcal{L}/\mathcal{L}_0$ at different $\gamma_d$ as a function of temperature with any other parameter fixed, for $u=2.0$ in Fig.\tref{fig:u2lor_vs_T} and $u=4.0$ in Fig.\tref{fig:u4lor_vs_T}. The violation for small and zero $\gamma_d$ can be interpreted considering again the structure of the transmission functions from the collection of real baths and probes in the energy window included into transport at each temperature. At small and zero dephasing, the sharp features of the transmission would prevent the Sommerfeld expansion necessary to directly derive the Wiedemann-Franz law from Eqs.\eqref{eq:e_current}-\eqref{eq:q_current} at low temperatures. As dephasing increases, however, these features are progressively broadened and the energy windows over which the transmissions are continuous gets larger, so the ratio $\mathcal{L}/\mathcal{L}_0$ is restored to $1$.

The fact that thermal and electrical conductivities can have maxima at different values of dephasing strength, translates to values of $\gamma_d$ where the magnitude of heat current is maximized at low corresponding magnitude of electric current or vice versa. In the next section we argue and demonstrate that this effect can be exploited in the context of steady-state thermal machines.

\section{Dephasing enhanced quasiperiodic machines}
\label{sec:frigo}
The set-up we study functions naturally as a thermoelectric device, with the Fibonacci chain acting as the working medium. We are free to regulate the thermodynamic parameters of the real baths, $T$, $\mu$ and biases $\Delta T$, $\Delta \mu$. We set $\Delta T >0$ and $\Delta \mu <0$. The electric and heat currents flowing from left to right is assumed to be the positive direction. By standard convention, the power 
\begin{align}
    P = J_e \Delta \mu
\end{align}
 is negative if it is extracted from a thermoelectric device, while it is positive if it is input into the thermoelectric device. If the temperature bias drives the electric current against the chemical potential difference, the electrons from the baths perform a certain amount of work per unit of time inside the central region, generating power. In this case we have
 \begin{equation}
    P <0,~~J_q>0~~\textrm{(heat engine regime).}
\end{equation}
The efficiency of the heat-to-work conversion is the same as for a standard cyclic thermal engine, given by
\begin{equation}
\label{eq:eff}
    \eta^{(h)} = \dfrac{-P}{J_q} \le \eta^{(h)}_C = 1 - \dfrac{T}{T + \Delta T},
\end{equation}
and it is bounded from above by the corresponding Carnot efficiency $\eta^{(h)}_C$. When, instead, the heat current is negative, as a consequence of the applied chemical potential difference, the machine acts as a refrigerator,
\begin{align}
    J_q < 0,~~P>0~~\textrm{(refrigerator regime)}.
\end{align}
In this case, heat is transported from the right (colder) to the left (hotter) bath, while power is supplied to the system ($P > 0$). The efficiency of the refrigeration is quantified by the coefficient of performance
\begin{equation}
\label{eq:cop}
    \eta^{(r)} = \dfrac{-J_q}{P} \le \eta_C^{(r)} = \dfrac{T}{\Delta T}.
\end{equation}
It is clear from the expressions for $\eta^{(h)}$ and $\eta^{(r)}$ that situations where the magnitude of heat current and the magnitude of electric current are maximized at different values of $\gamma_d$ will be advantageous if either of the currents is negative. 

There can be a third working regime of a two-terminal device, where both heat current and power are positive, $J_q>0$, $P>0$. In this so called accelerator regime, the electrical power input into the system heats up the two reservoirs. This is usually the most easily obtained regime, without much fine-tuning of parameters. Here, we will not be interested in this regime.

In linear response, the efficiency or performance of a two-terminal device maximized over the driving forces can be analytically expressed through a single dimensionless figure of merit $ZT$\tcite{goldsmid2010intro, benenti2017}
\begin{equation}
    \dfrac{\eta^{(h/r)}_{\text{max}}}{\eta^{(h/r)}_C}  = \dfrac{\sqrt{ZT +1} -1}{\sqrt{ZT+1} + 1},
\end{equation}
with
\begin{equation}
ZT = \dfrac{GS^2T}{K} = \dfrac{\sigma S^2T}{\kappa}.
\end{equation}
Larger values of $ZT$ correspond to higher efficiency or performance, giving the maximum theoretical limits for $ZT \rightarrow \infty$. It is also intuitive from above result that if the Wiedemann-Franz law is violated such that $K/(GT)<\mathcal{L}_0$, as we see in our case, it may aid the performance of the heat-engine or the refrigerator.

Thermoelectric response in nanoscale devices is linked to their energy-filtering properties\tcite{MahanS1996, whitney2014most,benenti2017}. If transport is blocked within a certain energy window, the Seebeck coefficient $S$ increases dramatically. This is generally achieved by tuning the thermodynamic variables of the reservoirs\tcite{jaliel2019, Popp2021} or by choosing samples that would exhibit strongly energy dependent transmission properties, for example in presence of a mobility edge\tcite{sivan1986, yamamoto2017, chiaracane}. 
%Thermoelectric effects also usually imply violations of the Wiedemann-Franz law, which would otherwise constrain the Lorenz number $\mathcal{L} = K/GT = \kappa/\sigma T$ to a constant.

\begin{figure}[t]
\centering
\subfloat[]{\includegraphics[width=0.5\columnwidth]{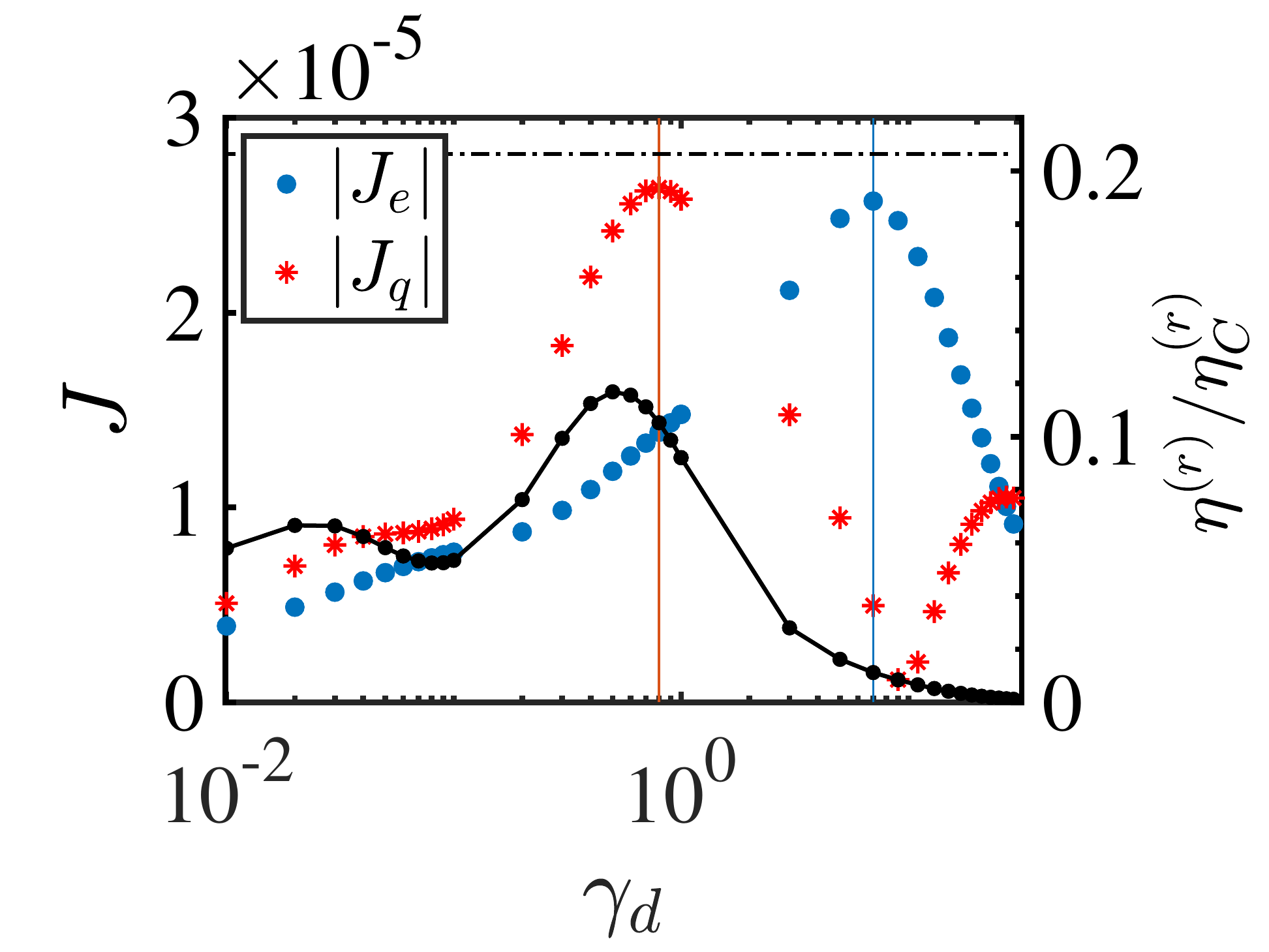}\label{fig:u4fridge1}}
\subfloat[]{\includegraphics[width=0.5\columnwidth]{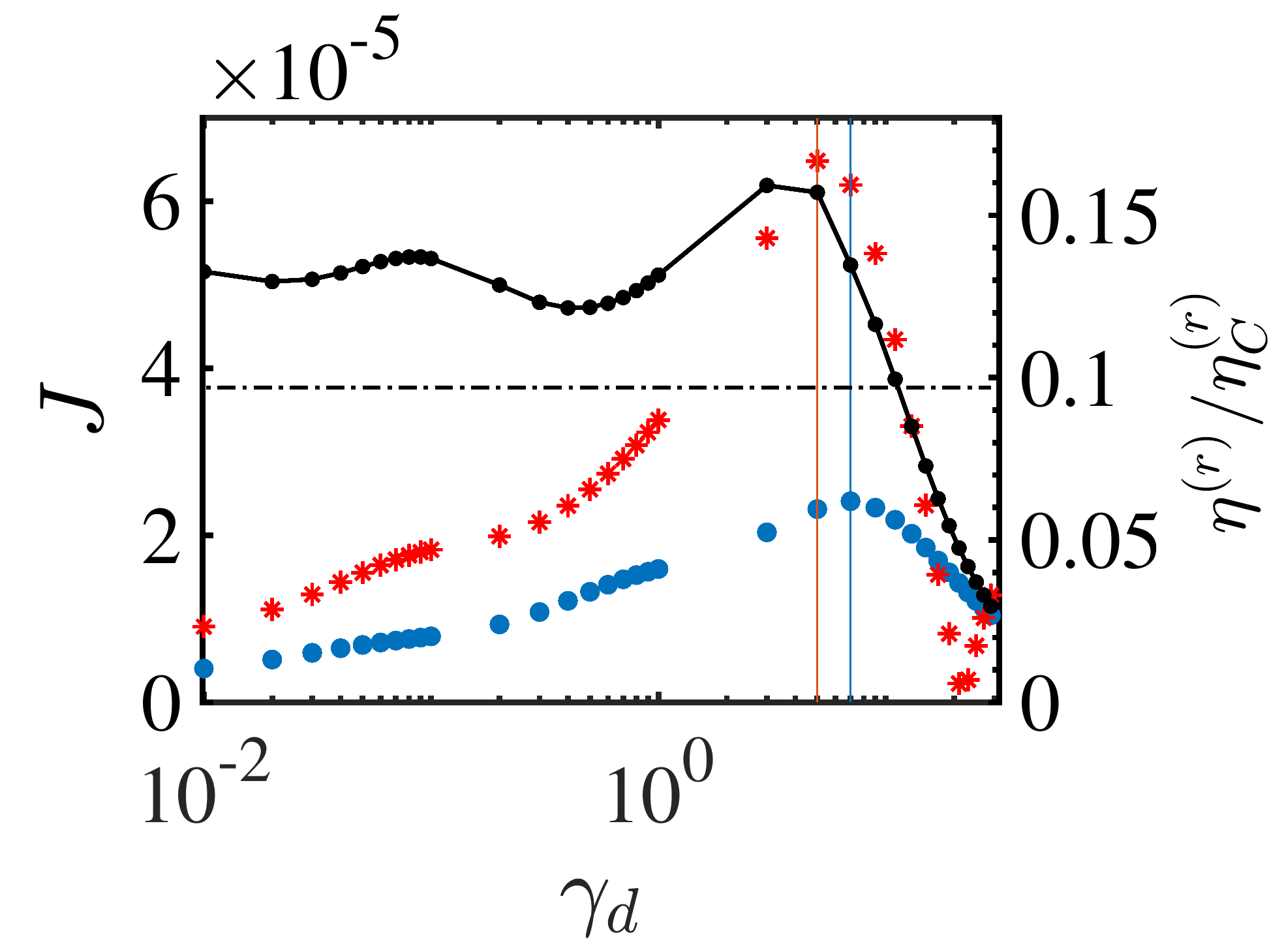}\label{fig:u4fridge2}} \\
\subfloat[]{\includegraphics[width=0.5\columnwidth]{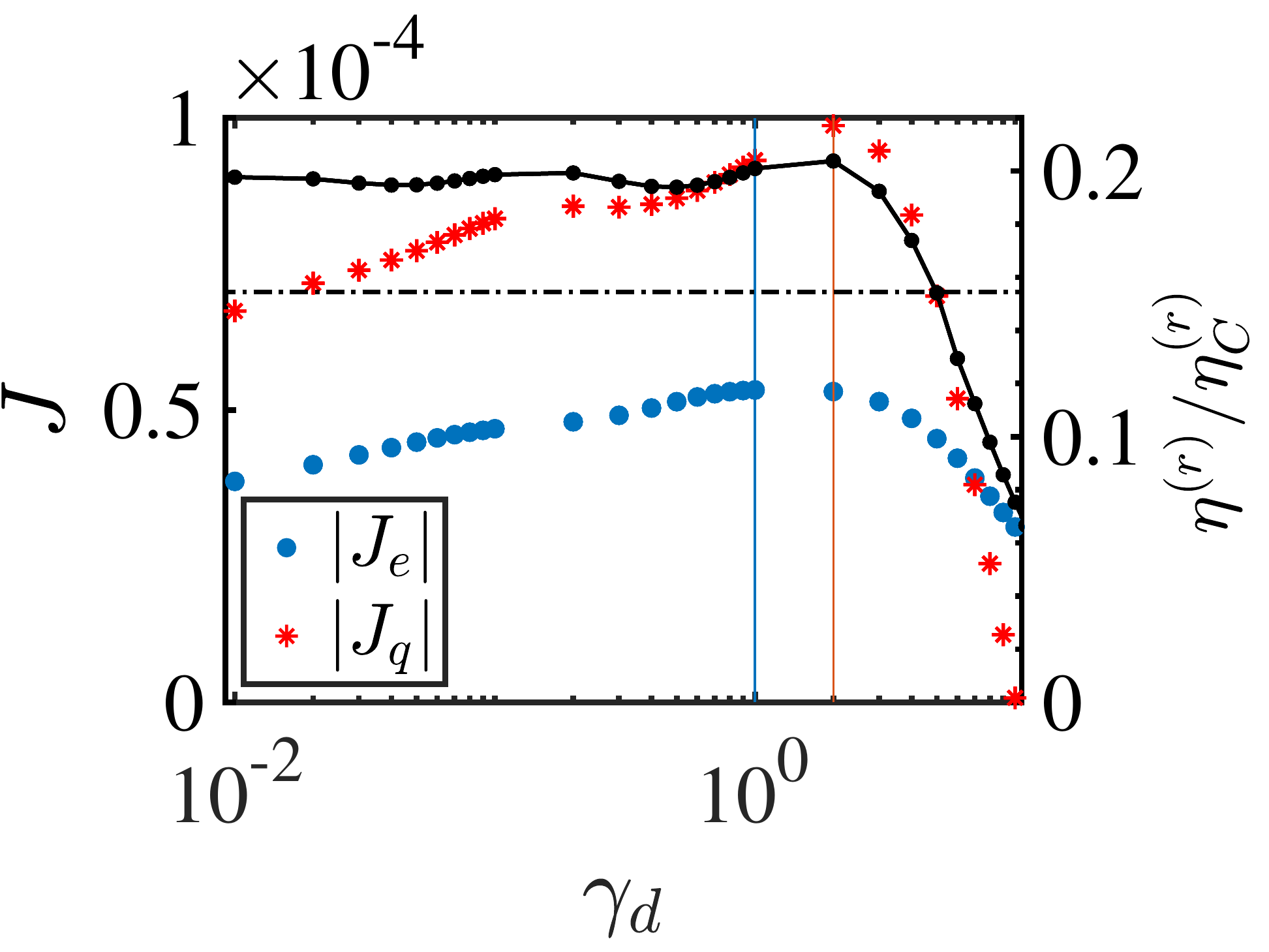}\label{fig:u2fridge1}} 
\subfloat[]{\includegraphics[width=0.5\columnwidth]{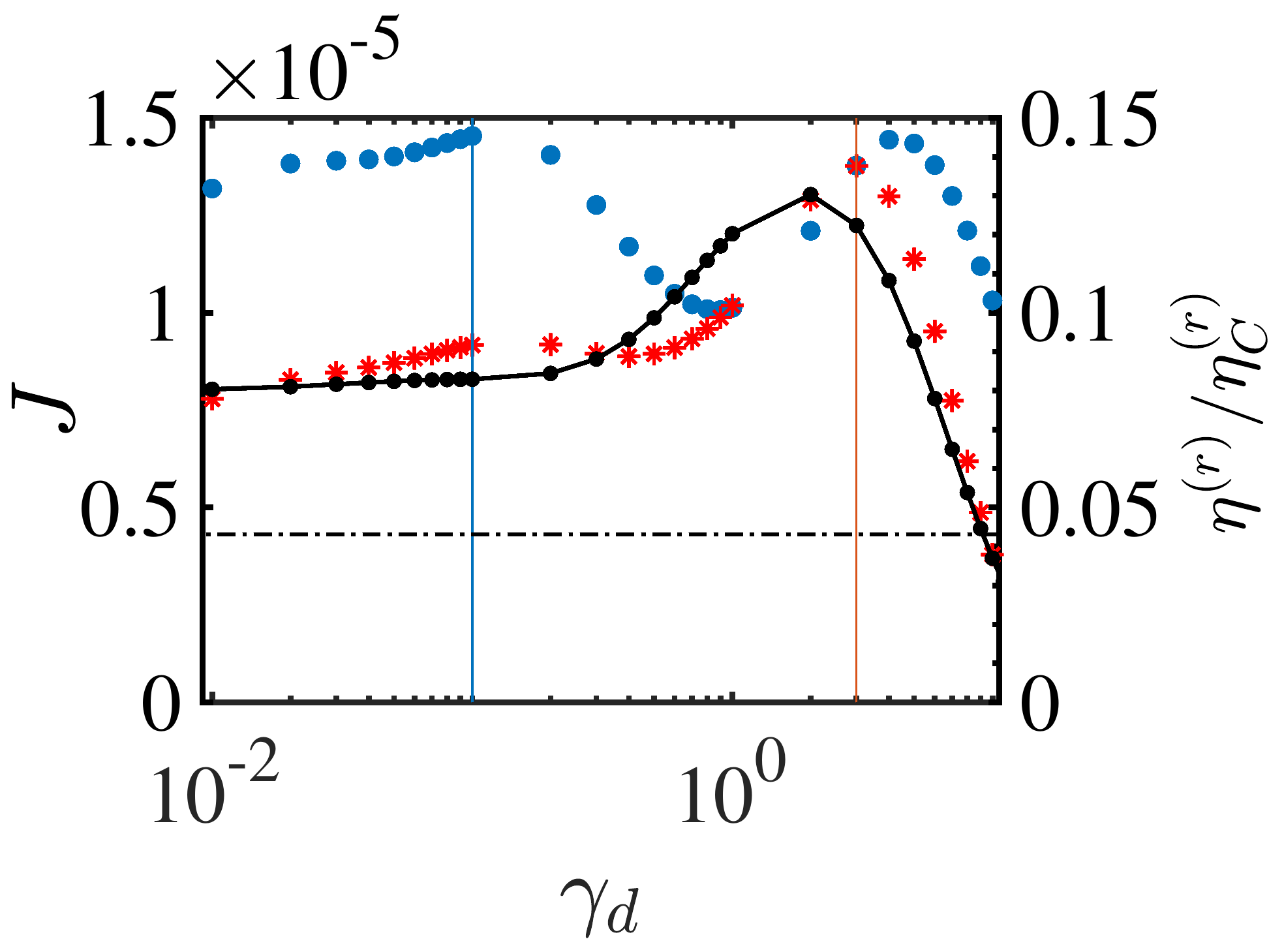}\label{fig:u2fridge2}} 
\caption{Examples of configurations which will function as a fridge, (a)-(b) for $u=4.0$ and (c)-(d) $u=2.0$, $N=200$. The red (blue) dots indicate the magnitude of the heat (electric) current, with its maximum highlighted by a vertical continuous line in the same colour. On the right axis, $\eta^{(r)}$ normalized to the maximum theoretical limit $\eta^{(r)}_C$ is shown in black, and its value at zero dephasing is indicated as a reference with a horizontal dashed line. Parameters: at $u=4.0$ (a)~$\mu = 0, T=5, \Delta T = 0.1, \Delta \mu = 0.5$, (b)~$\mu = -5.2, T=10, \Delta T = 0.5, \Delta \mu = -1.$, at $u=2.0$ (c)~$\mu = -3.3, T=10, \Delta T = 1.0, \Delta \mu = -1.0$, (d)~$\mu = 2.8, T=1.0, \Delta T = 0.01, \Delta \mu = 0.1$.}
\label{fig:fridge}
\end{figure} 

 To make a two-terminal device act as either a heat engine or a refrigerator, in absence of dephasing, it can be shown that a key ingredient is asymmetry of the transmission function around the chosen chemical potential \tcite{benenti2017}. The peculiar transmission function of the Fibonacci model in absence of dephasing, which reflects its fractal spectrum (see Fig.\tref{fig:transmission}), shows that it naturally has this property for various choices of chemical potentials, and thus can serve as working medium for a natural refrigerator or heat engine. Introducing incoherent inelastic scattering into the system makes it difficult to extrapolate the energy-filtering properties of the effective spectrum, since it is given by the collective transmissions of the fictitious probes. However, we have already seen that the non-trivial spectral properties of the original model make the conductivities highly sensitive to the dephasing strength, suggesting that particular thermodynamic configurations could realize efficient thermoelectric devices. 

A particularly interesting case occurs for parameters where the Fibonacci model in absence of dephasing is subdiffusive and works as either a refrigerator or a heat engine. As we have seen in previous sections, dephasing will increase the currents in this case, making transport diffusive. If the system still acts as a refrigerator (heat engine) it will therefore enhance its cooling rate $-J_q$ (power output $-P$). Moreover, if the maxima of electrical and heat currents are different, it can even increase the coefficient of performance (efficiency) of the refrigerator (heat engine). In the following, we demonstrate such simultaneous dephasing-induced enhancement of both cooling rate and coefficient of performance in the refrigerating regime.

We first scan the parameter space and select configurations which function as refrigerator. The plots in Fig.\tref{fig:fridge} show the absolute values of the electric (blue) and heat (red) currents as a function of $\gamma_d$ at different thermodynamic parameters. On the right axis, we also show $\eta^{(r)}/\eta^{(r)}_C$, whose value at zero dephasing is indicated by a dashed horizontal line. We observe explicitly in Fig.\tref{fig:u4fridge1} that electrical and heat currents have maxima at different values of dephasing strength. By definition, the coefficient of performance $\eta^{(r)}$ is maximized when the magnitude of heat current is maximum, but the electrical current is away from its maximum. However, $\eta^{(r)}$ for this choice of chemical potentials, temperatures and Fibonacci potential strength ($u=4.0$), is always below the value obtained in absence of dephasing.  In Fig.\tref{fig:u4fridge2}, instead, which shows a refrigerating regime for a different choice of chemical potentials and temperatures at the same value of $u$, we see $\eta^{(r)}$ enhanced by dephasing for a wide range of $\gamma_d$. For $u=2.0$, we can also find different configurations in Figs.\tref{fig:u2fridge1}-\tref{fig:u2fridge2} where the performance is enhanced by the presence of dephasing. 

Since the chosen values of the potential $u$ lie in the subdiffusive regime of the Fibonacci model at $\gamma_d = 0$ (see Fig.\tref{fig:GKvari}), the presence of inelastic scattering increases the currents by several orders of magnitude (see, for example Figs.\tref{fig:je_u4},\tref{fig:jq_u4}) and, consequently, dramatically enhances the cooling rate of these refrigerating regimes. Moreover, in Figs.\tref{fig:u4fridge2}-\tref{fig:u2fridge2}, we see even the coefficient of performance enhanced by the different sensitivity of the currents to dephasing strength.

\section{Conclusion}
\label{sec:concl}
We have studied the linear-response transport and thermodynamics of the Fibonacci chain both in the absence and the presence of dephasing noise from incoherent inelastic scattering at finite temperature. Specifically, we describe bulk inelastic scattering using the method of voltage-temperature probes within the Landauer-B\"uttiker framework of quantum transport. In absence of dephasing, the Fibonacci model shows anomalous transport which continously varies from superdiffusive to subdiffusive as a function of the Fibonacci potential strength. This fact was previously known in the limit of infinite temperature for particle or spin transport \tcite{varma2019, lacerda2021, chiaracane2}. We demonstrate that this fact survives at finite temperatures, and is observable in both electric and thermal transport, even in the presence of both temperature and chemical potential biases. We find that dephasing due to inelastic scattering makes both electric and thermal transport diffusive for all values of Fibonacci potential strength. This means that, in the parameter regime where the coherent model is subdiffusive, dephasing enhances both electrical and thermal transport. For diffusive transport, electric and thermal conductivities are well-defined and finite, allowing us to study them as a function of dephasing strength. We find that, in the regime where the coherent model is subdiffusive, at finite temperatures, the conductivities can show a non-monotonic behavior with increase in dephasing strength.  This is consistent with observations in previous works investigating spin transport by modelling dissipation and dephasing via Lindblad equations \cite{znidaric2017,lacerda2021}. However, surprisingly, at low and intermediate temperatures, we find occurrence of several chemical potential dependent local maxima in the conductivities as a function of the dephasing strength. Moreover, remarkably, we find a clear violation of Wiedemann-Franz law over a wide range of dephasing strength even at low temperatures, even though the transport becomes diffusive.  Further,  the optimal dephasing strength corresponding to the global maximum differs for the thermal and electric conductivities and is highly sensitive to the thermodynamic affinities. 
%We attribute all these surprising properties of the system to the fractal nature of the Fibonacci spectrum. 

One might expect that this highly non-trivial transport behavior is associated with the fractal structure of the Fibonacci spectrum, and we conjecture that this is indeed the case. However, it is challenging to find a more microscopic explanation in presence of dephasing at finite temperature. In the case of coherent transport, the transmission function for scattering processes connects the microscopic details of the system to the thermoelectric properties of the non-equilibrium steady state. Conversely, when dephasing is introduced through the probes, transport is determined by the entire collection of transmission functions between reservoirs and probes. This complexity makes the interplay between spectral properties, dephasing, and transport difficult to understand intuitively.

Nevertheless, our numerical results clearly indicate that thermal and particle transport behave differently with respect to dephasing. This opens the possibility of enhancing thermoelectric effects by noise. In particular, we have demonstrated a remarkable dephasing-induced-enhancement of both cooling rate and coefficient of performance simultaneously for autonomous refrigeration using the Fibonacci quasicrystal as a working medium. Although this finding is specific to certain parameter regimes of the Fibonacci model, we hope that the results might serve more generally as a conceptual guide for the realization of new synthetic systems for nanoscale heat management based on quasiperiodic potentials.

\textbf{Acknowledgements.} We thank G. Haack, R. Sánchez, and M. Žnidarič for the useful comments on the manuscript. This work was funded by the European Research Council Starting Grant ODYSSEY (Grant Agreement No. 758403) and the EPSRC-SFI joint project QuamNESS. J.~G. is supported by a SFI-Royal Society University Research Fellowship. A.~P. acknowledges funding from European
Unions Horizon 2020 research and innovation program under
the H2020 Marie Sklodowska Curie Actions Grant Agreement No. 890884. We acknowledge the Irish Centre for
High End Computing (ICHEC) for the provision of computational facilities.

\bibliographystyle{apsrev4-1}
\bibliography{fib_lbprobes}

\end{document}